%% file: sn-article.tex
\documentclass[pdflatex,sn-mathphys-num]{sn-jnl}

\usepackage{graphicx}%
\usepackage{multirow}%
\usepackage{amsmath,amssymb,amsfonts}%
\usepackage{amsthm}%
\usepackage{mathrsfs}%
\usepackage[title]{appendix}%
\usepackage{xcolor}%
\usepackage{textcomp}%
\usepackage{manyfoot}%
\usepackage{booktabs}%
\usepackage{algorithm}%
\usepackage{algorithmicx}%
\usepackage{algpseudocode}%
\usepackage{listings}%

\usepackage[english]{babel}
\usepackage{csquotes}
\usepackage{bm}
\usepackage{amsfonts}
\usepackage{comment}
\usepackage{soul,color}
\usepackage{subcaption}

\theoremstyle{thmstyleone}%
\theoremstyle{thmstyletwo}%

\theoremstyle{thmstylethree}%

\begin{document}

\title[Article Title]{Curved Mesh Adaptation for High-Order Finite
Element Simulations}

\author[1]{\fnm{Aditya Yogesh} \sur{Joshi}}\email{joshia5@rpi.edu}

\author*[1]{\fnm{Mark S.} \sur{Shephard}}\email{shephard@rpi.edu}

\affil[1]{\orgdiv{Scientific Computation Research Center; Department of Mechanical, Aerospace and Nuclear Engineering}, \orgname{Rensselaer Polytechnic Institute}, \orgaddress{\street{110 8th St.}, \city{Troy}, \postcode{12180}, \state{NY}, \country{USA}}}


\abstract{The ability to take advantage of computationally efficient high-order finite element methods to perform adaptive finite element analysis of complex engineering problems over general 3D domains requires the ability to adapt meshes with curved elements that maintain the level of geometric approximation of the domain boundary required. This paper presents a conforming curved mesh adaptation procedure aimed at effectively supporting automated adaptive analysis of problems for which the domain geometry is defined in a CAD system. The local mesh modification procedures interact with the CAD geometry to curve the mesh edges and faces representing those boundaries to the order of approximation of the high-order elements being used. The curved mesh edges and faces representing the domain boundaries are based on B\'ezier approximation geometry, which provides more accurate evaluations of surface-related quantities of interest than the commonly used interpolation methods.
To attain computational efficiency during mesh adaptation, the interior mesh entities have their geometric order kept as low as possible while still maintaining control of element shapes. The order of curved mesh entities is limited to no higher than cubic, which allows the definition of an effective procedure to define the shape of the limited number of interior mesh entities that must be curved. The procedures, which are fully parallelized, are demonstrated on the adaptive radio-frequency analysis of a magnetically confined fusion system containing a fully represented radio-frequency antenna.}

\keywords{Curved meshes, mesh adaptation, mesh modification}



\maketitle

\include{Introduction}

\include{MeshModelRepresentation}

\include{MeshEntitiesOnModelBdry}

\include{ShapeMeasure}

\include{CurvedMeshAdaptOverview}

\include{CoreMeshAdaptOperators}

\include{ElementShapeImprove}

\include{ExecuteCurvedMeshAdapt}

\include{Closing}

\bmhead{Acknowledgements}

This research was supported by the U.S. Department of Energy, Office of
Science, under awards DE-SC0021285 (FASTMath SciDAC Institute) and DE-SC0024369 (Center for Advanced Simulation of RF - Plasma - Material Interactions).
Any opinions, findings, and conclusions or recommendations expressed in this material are those of the author(s) 
and do not necessarily reflect the views of the U.S. Department of Energy.


\bibliography{references}

\end{document}

%% file: Introduction.tex
\section{Introduction}\label{Introduction}

High-order finite element methods demonstrate advantages over low-order finite element methods due to their higher rates of convergence and computational efficiency on today’s GPU-accelerated hardware~\cite{henneking2025real, andrej2024high}.
Although the highest levels of computational efficiency are attained using tensor product–based elements, hexahedra in 3D and quadrilaterals in 2D, the goal of this work is to provide effective, automated, and adaptive simulation capabilities for problems involving geometrically complex engineered systems.
The most direct way to achieve this goal is to take full advantage of the ability of existing simplex-element mesh generators to automatically generate meshes for geometrically complex engineering systems defined in CAD systems, and to adapt those meshes to ensure the desired level of accuracy.

A popular approach for conforming mesh adaptation is the use of cavity-based mesh modification operations that locally modify the mesh to match the prescribed mesh size field. A number of investigators have developed such procedures that can be applied to meshes that include curved elements. The majority of the procedures developed incorporated local mesh modification operations based on straight-sided elements to change elements distributions to meet a given mesh metric field followed by mesh entity curving operations applied in a manner to satisfy mesh properties desired for their application requirements~\cite{coppeans2025anisotropic, jiang2021bijectivenyu, Li2003a, lu2014parallel, rochery2021p2, loseille_opt_cavity2023, shi2024local}.

Curved mesh representations employed include standard interpolating polynomials~\cite{jiang2021bijectivenyu, rochery2021p2, Persson2009, shi2024local, coppeans2025anisotropic} or interpolating B\'ezier polynomials~\cite{loseille_measure, Li2003a, lu2014parallel}. In most references, care is taken to use optimal spacing of the interpolating points (Gauss-Labatto or Babuška-Chen). In the current paper, consideration is given to the potential advantages of using approximating B\'ezier polynomials for the shape of mesh edges and faces on curved domain boundaries.  

The methods to define the shape of the curved elements include the application of mesh shape optimization methods~\cite{loseille_measure, shi2024local}, typically applied at a local cavity level, or specific geometric property satisfaction, such as the quadratic edge fitting procedure presented in reference~\cite{zhang2021generationremacle}, and the cubic edge definition procedure presented in this paper. 

Maintaining convergence of the finite element method when the analysis geometric model includes curved surfaces requires that the mesh entities representing those surfaces be curved to the appropriate level of geometric approximation. When considering general domains, along with the size and distribution of elements dictated by an adaptive analysis procedure, it becomes clear that some of the mesh entities in the interior of the domain must also be curved in a way that provides good mesh quality.
The curved mesh adaptation procedures presented in this paper build on our previous work on the development of cavity-based mesh modification procedures for linear simplex~\cite{Li2003, Li2003a} and curved simplex~\cite{lu2014parallel} meshes.

The ability to support automated adaptive simulations requires employing a general representation of the analysis geometric model and the mesh, as well as supporting the maintenance of the relationship between mesh entities and the analysis geometric domain. Section~\ref{MeshRepresentation} indicates that the use of a boundary representation for the analysis geometric model and the mesh provides the ability to interact with analysis geometric models defined in CAD systems and with curved meshes. Section~\ref{MeshEntitiesOnModelBdry} discusses and compares the use of interpolation and geometric approximation for representing curved mesh entities on the domain boundary. The method used to determine invalid elements and the element shape measure are discussed in Section~\ref{ShapeMeasure}. Section~\ref{AdaptOverview} outlines the goals of the mesh adaptation procedures and provides an overview of the developed approach. Section \ref{CoreOperators} describes core curved mesh entity operations. The mesh adaptation operators are described in Section~\ref{ModificationOperators}, while Section~\ref{ImproveShape} presents the procedure used to improve element shapes. Section~\ref{ExecuteResults} demonstrates the application of curved mesh modification procedures, outlines the execution of the resulting adaptive procedures in RF simulations, and provides an example of the application on a fusion energy system.

%% file: MeshModelRepresentation.tex
\section{Mesh and Analysis Geometry Model Representations} \label{MeshRepresentation}
\subsection{Mesh and Model Topology}
Employing a boundary representation for both the analysis geometric model and the mesh supports the effective application of mesh adaptation through local mesh modification. A boundary representation consists of two components. The first is a graph in which the nodes represent topological entities and the edges represent selected adjacencies between those entities. The second component is the geometric shape information associated with the appropriate topological entities in the boundary representation.

In the case of mesh topology, it is assumed that any mesh region (3D volume element) is bounded by a single shell of adjacent mesh faces, and each mesh face is bounded by a single loop of adjacent mesh edges. Under this assumption, a complete mesh boundary representation need only include the primary boundary entities, regions, faces, edges, and vertices, and a set of mesh adjacencies to define a complete representation. A complete representation is one in which any of the 12 possible first-order mesh adjacency operations can be evaluated in $O(1)$ time. Reference~\cite{beall1997general} discusses options for complete mesh topological representations. The mesh modification operators presented in this paper employ the one-level adjacency structure~\cite{beall1997general, ibanez2016pumi, ibanez2016conformal}, which maintains both the upward and downward adjacencies between regions and faces, faces and edges, and edges and vertices.
The shape information associated with the mesh entities is discussed in the next subsection.

The analysis CAD models can have regions with multiple shells and faces with multiple loops.
The analysis geometric model boundary representation includes regions, shells, faces, loops, edges and vertices. Since the analysis geometric models to be supported can be general combinations of different material regions and reduced dimension model entities (e.g., dangling faces not bounding any region), a general non-manifold topological model representation is required~\cite{weiler1985edge, Weiler1986}. 
In this work it is assumed the analysis model will be defined in some geometric modeling software such as Parasolid~\cite{Parasolid}, ACIS~\cite{ ACIS} or OpenCascade~\cite{Open_CASCADE}. 
The shape information associated with the model faces and edges in CAD systems is most commonly Non-Uniform Rational B-Splines (NURBS) defined in a parametric coordinate system that is mapped to a global Cartesian coordinate system. However, alternative representations of the geometric shape information are possible, for example in one ongoing fusion system modeling project, the geometry is defined in terms of Fourier coefficients. Since the geometric modeling systems support Boolean operations on geometric model components, surfaces can be trimmed and regions cannot be defined in a simple parametric coordinate system. On the other hand, the geometric modeling kernels support application programming interface operations that allow programmers to execute specific geometric queries. The mesh adaptation procedures presented in this paper employ the geometric modeling system API to interact with analysis geometric model shape information~\cite{beall2003accessing}. 

\begin{figure}
\centering
\captionsetup{width=\linewidth, justification=centering}
    \includegraphics[width=.55\linewidth]{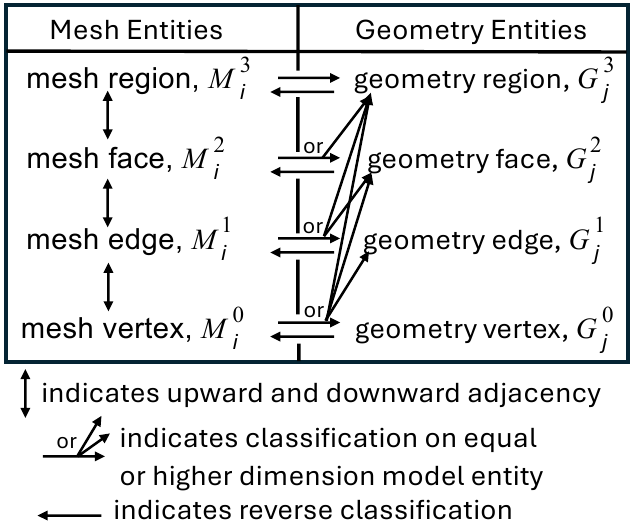}
    \caption{Mesh topology and relationship of geometric model and mesh entities.}
    \label{fig:classification}
\end{figure}

For the purpose of providing a technical description of the curved mesh adaptation operations presented in this paper, it is useful to employ a concise notation for the mesh and geometric model entities~\cite{beall1997general}. Figure~\ref{fig:classification} illustrates the mesh entities and, through the use of vertical double arrows, indicates a one-level adjacency structure. This structure provides a complete representation that can be evaluated in O(1) time.

To support the ability to relate mesh topological entities to the primary topological entities in the analysis geometric model, the concept of mesh classification is introduced. Each mesh entity is uniquely classified with respect to a geometric model entity of equal or higher dimension. This relationship can be concisely stated as follows:
\begin{equation}
    M_i^{d_i} \sqsubset G_j^{d_j}
\end{equation}
where $M_i^{d_i}$ is the $i^{th}$ mesh entity of dimension $d_i$, $G_j^{d_j}$ is the $j^{th}$ geometric model entity of dimension $d_j$  and $d_j \ge d_i$.
As indicated by the right pointing and upward angled arrows in Figure~\ref{fig:classification}, a mesh region, $M_i^3$, is classified as being within a geometric model region, $G_j^3$. A mesh face, $M_i^2$, is classified as being within a geometric model region, $G_j^3$, or on a geometric model face $G_j^2$.  A mesh edge, $M_i^1$, is classified as being within a geometric model region, $G_j^3$, on a geometric model face $G_j^2$, or on a geometric model edge $G_j^1$. A mesh vertex, $M_i^0$, is classified within a geometric model region, $G_j^3$, on a geometric model face $G_j^2$, on a geometric model edge $G_j^1$, or at a geometric model vertex $G_j^0$.  

In addition to mesh entity classification, a relationship between geometric model topological entities and mesh topological entities is employed during curved mesh adaptation. This relationship, known as reverse classification, provides the set of equal-order mesh entities that are classified on a given model entity.
This is concisely stated as
\begin{equation}
    \{M_i^{d_i}\} \sqsupset G_j^{d_i}
\end{equation}
where $\{M_i^{d_i}\}$ is the set of  mesh entities of dimension $d_i$ classified on model entity $G_j^{d_i}$. 
The two reverse classifications used in curved mesh adaptation are: $\{M_i^{1}\} \sqsupset G_j^1$ which is the set of mesh edges classified on model edge $G_j^{1}$ and  $\{M_i^{2}\} \sqsupset G_j^2$ which is the set of mesh faces classified on model face $G_j^{2}$. The left facing arrows on figure~\ref{fig:classification} indicate reverse classification. 

\subsection{Geometric Representation of Curved Mesh Entities} \label{sec:repre-bezier-form}
A common form of geometric shape information used to define curved mesh edges, faces and regions is interpolation. An alternative mesh entity representation used in this, and other works~\cite{luo2004automatic, Luo2002, rochery2023fast},  is B\'ezier polynomials where computational advantage can be taken of their convex hull property, efficient degree elevation and subdivision algorithms, and the fact that derivatives and products of B\'ezier functions are
easily computed B\'ezier functions~\cite{bezier1986, Farin1992}.

An order $n$ B\'ezier curve used to represent a mesh edge $M_j^1$ can be written as
\begin{eqnarray}
\textbf{X}^{(n)}(M_j^1) = 
\sum_{i=0}^n{b}_i^{(n)}(\xi)\textbf{C}_i^{(n)}; \hspace{5pt} \xi\in[0,1]
\label{eq:bezcurve}
\end{eqnarray}
where ${b}_i^{(n)}(\xi)$ is the $i^{th}$ Bernstein basis function of
degree $n$ defined as
\begin{equation}
b_i^{(n)}(\xi)=\binom{n}{i} \xi^i (1-\xi)^{n-i}; \hspace{5pt} i=0, 1, ..., n.
\label{eq:bern}
\end{equation}
A single-parameter Bernstein-B\'ezier polynomial maintains mapping between the curve representation in the parametric space and the physical space.
If the control points are in $3D$, i.e. 
$\textbf{C}_i = (C_x, C_y, C_z)^T \in R^3 $, then $\textbf{X}(\xi)$ maps a one-dimensional parametric coordinate system ($\xi\in R$) to a $3D$ Cartesian coordinate system ($\textbf{X}(\xi)\in R^3 $).

The $n^{th}$ order B\'ezier  representation of a high-order triangular mesh face $M_j^2$ can be written as
\begin{equation}
\label{eq:bez-tri}
\textbf{X}^{(n)}(M_j^2) = 
\sum_{i=1}^q b^{(n)}_{i} (\bm{\xi}) \textbf{C}^{(n)}_{i}
\end{equation}
where $b^{(n)}_{i}(\bm\xi)$ are the order $n$ Bernstein polynomials for parameter $\bm\xi = \{ (\xi_1,\xi_2,\xi_3) \;\; | \;\; \xi_1+\xi_2+\xi_3 = 1$ \& $(\xi_1,\xi_2,\xi_3) \in [0,1] \} $.
Note that one parametric coordinate, say $\xi_3$, is a dependent coordinate in that $\xi_3 = 1- (\xi_1+\xi_2)$,
and $q$ is the total number of control points.
If the face control points are in $3D$, i.e. 
$\textbf{C}_i = (C_x, C_y, C_z)^T \in R^3 $, then $\textbf{X}(\xi)$ maps a two-dimensional parametric coordinate system ($\bm\xi\in R^2$) to a $3D$ Cartesian coordinate system ($\textbf{X}(\xi)\in R^3 $), and vice versa.

The $n^{th}$ order B\'ezier based volume representation of high-order tetrahedron $M_j^3$ can be written as
\begin{equation}
\label{eq:bez-tet}
\textbf{X}^{(n)}(M_j^3) = 
\sum_{i=1}^q b^{(n)}_{i} (\bm{\xi}) \textbf{C}^{(n)}_{i}
\end{equation}
where $b^{(n)}_{i}(\bm\xi)$ are the order $n$ Bernstein polynomials for parameter $\bm\xi = \{ (\xi_1,\xi_2,\xi_3,\xi_4) \;\; | \;\; \xi_1+\xi_2+\xi_3+\xi_4 = 1$ \& $(\xi_1,\xi_2,\xi_3,\xi_4) \in [0,1] \} $.
Note that one parametric coordinate, say $\xi_4$, is a dependent coordinate in that $\xi_4 = 1- (\xi_1+\xi_2+\xi_3)$.
For the tetrahedron, there are
3 components to control points $C_i^{(n)}$ in each direction 
$\textbf{C}_i = (C_x, C_y, C_z)^T \in R^3$,
and $q$ is the total number of control points for the $n^{th}$ order B\'ezier tetrahedron in $3D$. 

Table~\ref{tab:npts} indicates the total number of control points and the number of interior control points, for the $n^{th}$ order mesh edge, triangular face and tetrahedral region. The importance of noting the number of interior control points is that, when hierarchically setting the shape of mesh entities from edges, to faces, to regions, the control points on the boundary of the face and region have already been determined.  

\begin{figure}[htbp]
\centering
\captionsetup{width=\linewidth, justification=centering}
  \setlength{\fboxrule}{0.5pt}\fbox{\includegraphics[width=0.98\textwidth]{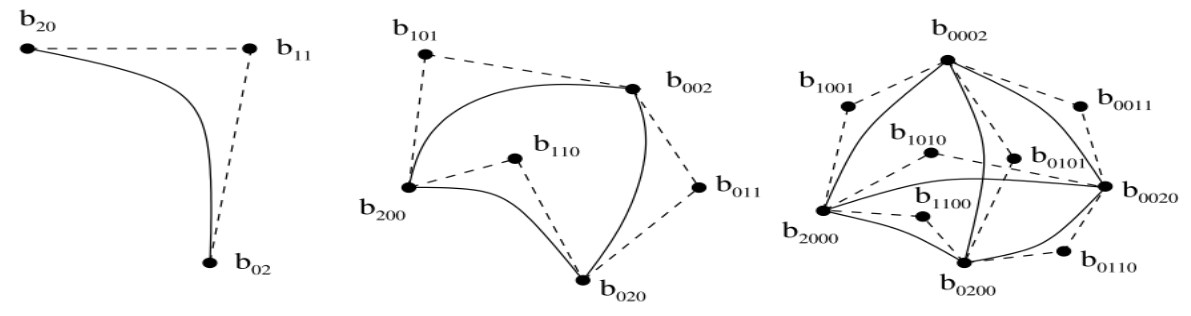}}
	 \caption{$2^{nd}$-order B\'ezier form of curved mesh edge $M_j^1$, face $M_j^2$, and region $M_j^3$.}
    \label{fig:2ndordertet}
\end{figure}

An illustration of the high-order point information for quadratic curved entities is shown in Figure~\ref{fig:2ndordertet}.
For the high-order curved triangle and tetrahedron, terms associated with the internal control points of the surface and volume are included in Eqs.~\ref{eq:bez-tri} and~\ref{eq:bez-tet}, as explained in \cite{Farin1992}.

\begin{table}
\captionsetup{width=\linewidth, justification=centering}
\caption{Formulations for the number of total and internal control points.}
\centering
\begin{tabular}{|c|c|c|}
\hline
Entity type & Total & Internal\\[2ex]
\hline\hline
  Curve & $n+1$ & $n-1$ \\[2ex]
 \hline
  Triangle & $\frac{1}{2}(n+1)(n+2)$ & $\frac{1}{2}(n-1)(n-2)$ \\[2ex]
 \hline
  Tetrahedron & $\frac{1}{6}(n+1)(n+2)(n+3)$ & $\frac{1}{6}(n-1)(n-2)(n-3)$ \\[2ex]
  \hline
\end{tabular}
\label{tab:npts}
\end{table}

%% file: MeshEntitiesOnModelBdry.tex
\section{Mesh Entity Representation of Model Boundaries} \label{MeshEntitiesOnModelBdry}

Although theory indicates that the geometric approximation order required for convergence in the energy norm is one order lower than the basis order of the finite elements used~\cite{ciarlet, oden2012introduction}, numerical studies of the convergence of selected quantities of interest indicate that better results are obtained when the geometric approximation is of the same order as the finite element basis being used.

CAD system kernels provide APIs that support pointwise geometric information queries. The primary geometric operation used in this work is to request the global coordinates of a point on a given model edge or face given its parametric value(s) with respect to the model edge or face. See Dey et al.~\cite{Dey2006, Dey1997a} for a discussion of the use of CAD and mesh entity parametric coordinates.

The next subsection briefly reviews the application of interpolation to define the shape of mesh edges and faces classified on curved model edges and faces. The subsequent subsection describes the least-squares fit method used to define the shape of curved mesh edges and faces representing portions of curved model boundaries. This section demonstrates that this approach, although more computationally expensive than interpolation,
 produces more accurate geometric fits, particularly when the mesh entity spans a substantial portion of a curved model entity, as is common in the target application of RF simulations that the current adaptive procedures support.

\subsection{Mesh Representation based on Interpolation}\label{sec-interp-geom-repre}

An important consideration in the application of interpolation to approximate a geometric shape is the selection of the locations of the interpolating points. Consider the most straightforward case of fitting a curve along an edge where the Jacobian of the mapping between the parametric coordinate values and the real coordinates is constant. In that case, the relative spacing of the interpolating points in both parametric and real space remains constant. It is known that even in this case, interpolating through uniformly placed points can produce noisy geometry and can reduce solution accuracy~\cite{chen1995approximate}. The use of optimal point placement, such as Gauss-Lobatto~\cite{Abramowitz1983Handbook} or Babuška-Chen points~\cite{chen1995approximate, bc_pts}, is less prone to noise and recovers the base finite element convergence rate.

When selecting locations to be interpolated on a CAD model edge, the most cost-effective approach is to apply the selected spacing (e.g., Gauss-Lobatto) in parametric space, since the only cost-effective CAD system API provides the real coordinates given the parametric coordinates on the model edge. The problem that arises is that the Jacobian of the mapping may vary along the model edge, in which case it is no longer clear how to define optimal point placement.
An even more serious situation occurs when defining an interpolating mesh edge classified on a model face. As discussed in~\cite{zhang2021generationremacle}, an edge with the shortest arc-length is desired. However, as demonstrated in the next subsection, the arc-length of the curve defined by interpolating using optimal point placement along a straight line in parametric space can yield mesh edges that are not close to the shortest arc-length, even for surfaces as common as a torus.

It should be noted that a procedure that attempts to select a more optimal curve on the surface by projecting a line onto the surface would need to employ the CAD system closest-point API, which operates by providing the coordinates of a point in space near a face and requesting the coordinates of the closest point on that face. The closest-point API is orders of magnitude more computationally demanding than the API that provides real coordinates given parametric coordinates. The next subsection presents an alternative approach to defining the geometry of curved mesh entities classified on curved model edges and faces. This approach is more expensive than interpolation using point placement based on model-entity parametric space, as it requires a substantial number of API calls to obtain real coordinates from parametric coordinates, but it is less expensive than procedures that employ closest-point API calls.

To demonstrate the advantages of the geometry approximation method presented in the next subsection over interpolation, the two approaches are compared for representing curved mesh entities classified on curved model geometries.

To avoid the need to support mesh modification operations for both interpolant-based and B\'ezier-based geometry, the mesh entity geometry determined by interpolation is converted to its B\'ezier representation using the inverse transformation, which requires solving a system of linear equations.

For instance a $4^{th}$ order B\'ezier curve is expanded as
\begin{equation}
\begin{aligned}
\textbf{B}^{(4)}(\xi_i)\bm{C} = (1-\xi_i)^4 \textbf{C}_0 + 4\xi_i(1-\xi_i)^3 \textbf{C}_1 + \\
6\xi_i^2(1-\xi_i)^2 \textbf{C}_2 + 4\xi_i^3(1-\xi_i) \textbf{C}_3 + \xi_i^4 \textbf{C}_4
\end{aligned}
\end{equation}

Given a set of nodal interpolation points $\xi_i, , i = 0,1,\ldots,4$, the B\'ezier curve is constructed to interpolate the locations $\textbf{L}_i$ that approximate the CAD model geometry.

\begin{equation}
\textbf{B}^{(4)}(\xi_i)\bm{C} = \textbf{L}_i, \quad i = 0,1,...,4
\end{equation}

This leads to a system of $5*d$ linear equations where $d$ is the spatial dimension. 

\begin{equation}
\begin{bmatrix}
(1-\xi_0)^4 & \cdots & \xi_0^4 \\
\vdots      & \ddots & \vdots \\
(1-\xi_4)^4 & \cdots & \xi_4^4
\end{bmatrix} \left[ \begin{array}{c} \textbf{C}_0 \\ \vdots \\ \textbf{C}_4 \end{array} \right]= \left[ \begin{array}{c} \textbf{L}_0 \\ \vdots \\ \textbf{L}_4 \end{array} \right]
\end{equation}

As a result, the 5 control points $\textbf{C}_i$ can be determined uniquely by solving the linear system.

\subsection{Mesh Representation based on Least Squares Fit} \label{sec:crv-fit}

In this work, a geometric approximation procedure is used to fit a B\'ezier representation to the model geometry using the method of least squares~\cite{borges2002total, Farin1992, pastva1998thesis}. For each mesh entity classified on a model boundary, the B\'ezier fit of the mesh edge or face is performed by executing the following procedure.

The order-$n$ B\'ezier curve is expanded.
\begin{equation}
\begin{aligned}
\bm{B}^{(n)}(\xi_i)\bm{C} = (1-\xi_i)^n \textbf{C}_0 + n\xi_i(1-\xi_i)^{n-1} \textbf{C}_1 \;+ \;..\\
.. \; + \; n\xi_i^{n-1}(1-\xi_i) \textbf{C}_{n-1} + \xi_i^n \textbf{C}_n
\end{aligned}
\end{equation}

To ensure the required $C^0$ interelement continuity, the boundaries of the curved mesh edges and faces must be common for each of the elements they bound. In the case of mesh edges, this is achieved by ensuring that every mesh vertex classified on the model boundary, ${M_i^0} \sqsubset G_j^k, k = 0, 1 \text{ or } 2$, is placed on the appropriate geometric model boundary entity.
With the mesh vertices set, the geometric approximation of the mesh edges classified on the model boundary, ${M_i^1} \sqsubset G_j^k, k = 1 \text{ or } 2$, is constructed using the least-squares fit procedure. This is followed by defining the mesh face geometric approximation, where the geometry of the face’s bounding edges has already been set.

Since the boundaries of the mesh edges and faces are set, the least-squares fit used to determine the B\'ezier mesh entity geometry only computes the locations of the interior control points. The least-squares fit of a mesh edge or mesh face is based on using $m$ data points $\textbf{D}_i$ generated on the model entity on which it is classified. These points are defined by sampling locations in the element domain, as dictated by parametric queries through the CAD kernel.
For an order $n$ curve, the control points $\bm{C}={\textbf{C}_i; i = 0,1,2, \ldots, n}$ are computed to minimize the distance to the boundary in the Frobenius norm $||\bm{B}(\xi)\bm{C} - \bm{D}||_F$, where the data points $\bm{D}$ are ordered such that $0 < \xi_1 < \xi_2 < \ldots < \xi_m < 1$ in the parametric space of the model entity on which it is classified.
This leads to a system of normal equations in matrix form
\begin{equation}
\begin{bmatrix}
(1-\xi_1)^n & \cdots & \xi_1^n \\
\vdots      & \ddots & \vdots \\
(1-\xi_m)^n & \cdots & \xi_m^n
\end{bmatrix} \left[ \begin{array}{c} \textbf{C}_0 \\ \vdots \\ \textbf{C}_n \end{array} \right] = \left[ \begin{array}{c} \textbf{D}_1 \\ \vdots \\ \textbf{D}_m \end{array} \right]
\end{equation}
written simply as
\begin{equation}
\bm{B}^{(n)}(\xi_i)\;\bm{C} = \textbf{D}_i
\end{equation}

The matrix $\bm D$ contains the co-ordinates of data points on the model and includes a column for each dimension in physical space, the matrix $\bm C$ contains the co-ordinates of control points, and the matrix $\bm B(\xi)$ is the Bernstein matrix.
Therefore, instead of $n+1$ points for order-$n$ edge used in the interpolation
process we use a larger dataset of $m$ number of points $m > n^2/4$ to get a more accurate representation \cite{dahlquist1974equidistant, pastva1998thesis}, and construct the B\'ezier curves that approximate the exact model geometry.
The approach for solving the system of equations discussed in \cite{Farin1992} is to multiply both sides with $\bm B^T$ to obtain a square and symmetric coefficient matrix $\bm B^T \bm B$ and solve:
\begin{equation}
\bm{B}^T\bm{B}\;\bm{C} = \bm{B}^T\textbf{D}
\end{equation}
The condition $m > n^2/4$ of having sufficient number of data points on the model ensures that the least squares approximation is well-conditioned as discussed by the reference \cite{dahlquist1974equidistant}.

\subsection{Comparison of Least Squares Fit and Interpolation for Representing Curved Model Boundaries} \label{CompareFit}

Although the meshes for accurate RF antenna and tokamak simulations can have several million elements, even when fifth-order finite elements are used, the size of individual elements on the smaller curved model entities is large compared to the size of the model entity. 
For instance, on typical initial meshes used in this work, there are often no more than four mesh edges on a cylindrical cross section. 
Therefore, when using a coarse initial mesh and curving entities on the boundary to the required level of approximation, it is important to quantify the accuracy of the geometric approximation.

\begin{figure}
\centering
\captionsetup{width=\linewidth, justification=centering}
	\includegraphics[width=0.3\textwidth]{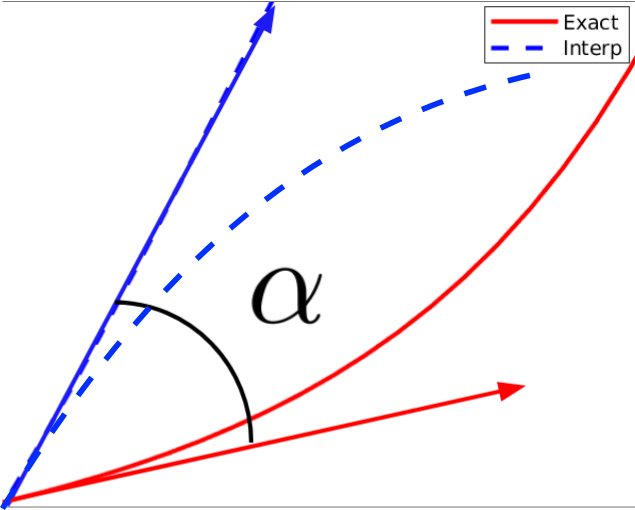}
	\caption{Example of angle measured at a point between the two curves.}
	\label{fig:angle}
\end{figure}

A number of parameters can be used to compare interpolation and curve fitting. The most common measures are the maximum and mean differences between the curves, percent error in curve edge length, and percent error in the area under the curve. In the case of the RF simulations of interest in this work, N\'ed\'elec elements that employ vector shape functions defined in terms of tangents are used. For these elements, the difference in tangent angles of the curves is an important parameter (see Figure~\ref{fig:angle}). Given this, the comparisons include measures related to the difference in tangent angles, $\alpha$, in terms of the maximum, mean, and integral of $\alpha$. Note that these measures are determined numerically by discretizing the curves into $P = 10^9$ equally spaced intervals, over which the maximum and mean are computed. The trapezoidal rule is used to estimate the integral of $\alpha$ over the curve.
\begin{equation}
    \int_{-1}^{1} \alpha\; dx = \sum_{k=1}^P \frac{\alpha(x_k) + \alpha(x_{k-1})}{2}\Delta x_k
    \label{eq:trapzIntg}
\end{equation}

\begin{figure}[htbp]
\centering
\captionsetup{width=\linewidth, justification=centering}
	\includegraphics[width=1.0\textwidth]{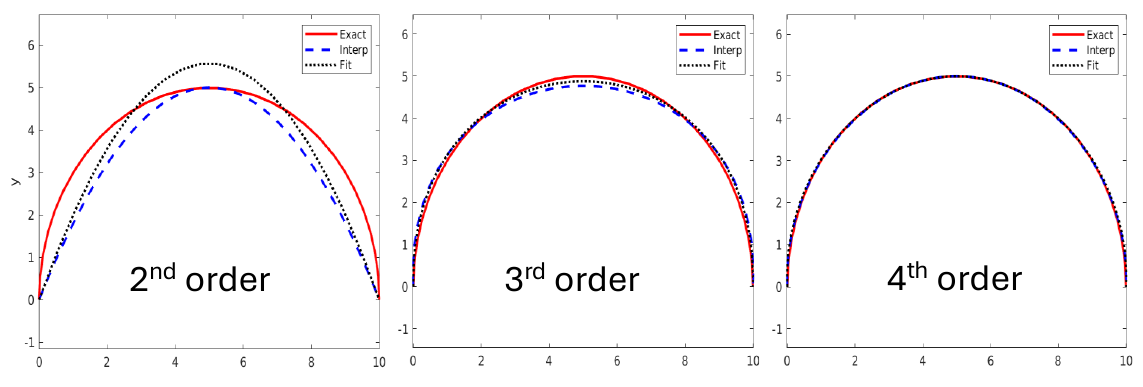}
	\caption{Comparison of interpolation and fitted geometry for a semicircle}
	\label{fig:Semi-circle}
\end{figure}

The first comparison of interpolation and curve fitting considers the approximation of a semicircle with a single mesh edge using second to fourth-order B\'ezier geometry. Table~\ref{tab:semicircle-fit} provides measures of the ability to approximate the semicircle, and Figure~\ref{fig:Semi-circle} shows a comparison of the curves. As can be seen for this case of a constant-curvature edge, both approximations converge well, with the fitted geometry typically being more accurate.

\begin{table}
\captionsetup{width=\linewidth, justification=centering}
\caption{Comparison of geometric interpolation and curve fitting for semicircle having exact length 15.70796 and exact area 39.26991.}
\centering
\begin{tabular}{|l|c|c|c|c|c|c|c|}
\hline
Order &Max $\alpha$&Integral $\alpha$&Mean $\alpha$& Mean Dist& Max Dist& \%Er Len& \%Er Area\\[1ex]
\hline\hline
2 Int. & 32.2& 72.5& 7.25& 0.351 & 0.669 & -5.85 & -15.1\\[1ex]\hline
2 Fit & 29.8& 72.2& 7.22& 0.350 & 0.575 & -2.11 & -5.36\\[1ex]\hline
3 Int. & 5.40& 34.0& 3.10& 0.131 & 0.233 & -1.28 & -1.08\\[1ex]\hline
3 Fit & 4.92& 30.9& 2.80& 0.128 & 0.227 & 0.505 & -0.175\\[1ex]\hline
4 Int. & 2.84& 3.34& 0.519& 0.020 & 0.035 & 0.398 & -2.61\\[1ex]\hline
4 Fit & 2.18& 3.09& 0.516& 0.018 & 0.031 & 0.206 & -2.58\\[1ex]\hline
\end{tabular}
\label{tab:semicircle-fit}
\end{table}

\begin{figure}[htbp]
\centering
\captionsetup{width=\linewidth, justification=centering}
	\includegraphics[width=1.0\textwidth]{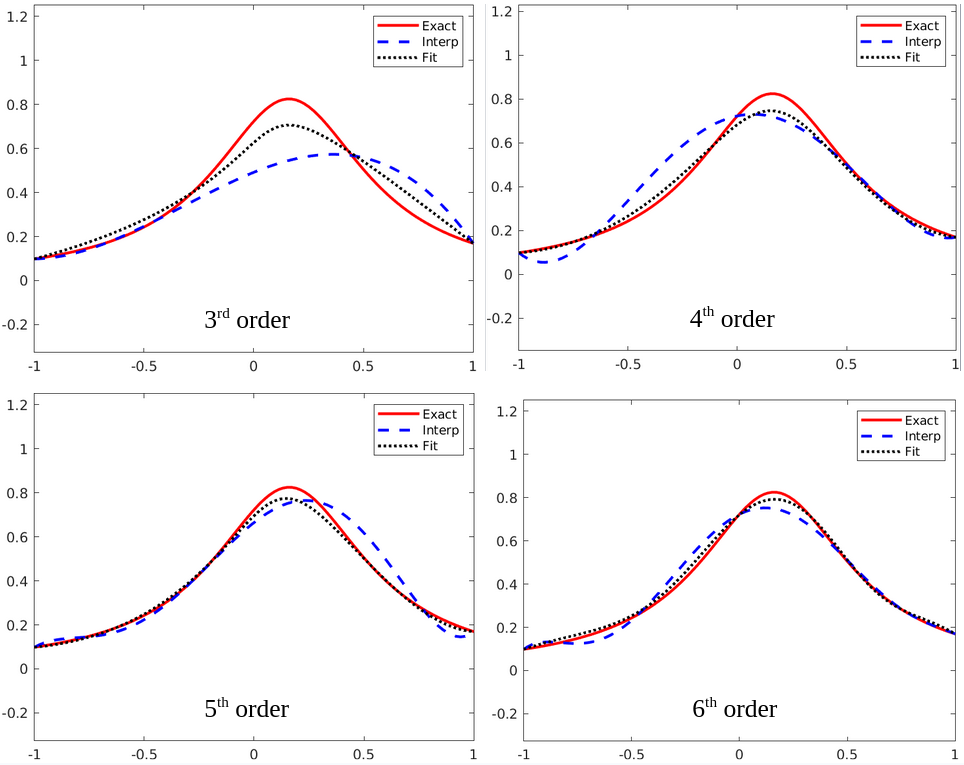}
	\caption{Curve approximation for the second test curve.}
	\label{fig:Curve-approximation}
\end{figure}

A second test case is asymmetric curved defined by 
\begin{equation}
    y = \frac{0.825}{1+5.5(0.16-x)^2} \hspace{10pt}; \hspace{10pt} x\in[-1,1]
\label{eq:rungekxsq_asym}
\end{equation}
Table~\ref{tab:runge-kxsq-asym} provides measures of the ability to approximate the asymmetric curve, and Figure~\ref{fig:Curve-approximation} shows a comparison of the curves. As can be seen from Table~\ref{tab:runge-kxsq-asym}, the fitted curve has much smaller differences in tangents than the interpolated curve in all cases.

\begin{table}
\captionsetup{width=\linewidth, justification=centering}
\caption{Comparison of geometric interpolation and curve fitting for the second example having exact length 2.50648 and exact area 0.81599.}
\centering
\begin{tabular}{|l|c|c|c|c|c|c|c|}
\hline
Order&Max $\alpha$&Integral $\alpha$&Mean $\alpha$&Mean Dist&Max Dist&$\%$Er Len&$\%$Er Area\\[1ex]\hline\hline
3 Int. & 51.7& 36.3& 18.2& 0.094& 0.284& -10.5& -8.76\\[1ex]\hline
3 Fit & 35.6& 22.0& 12.1& 0.030& 0.133& -6.91& -4.28\\[1ex]\hline
4 Int. & 47.4& 29.9& 14.9& 0.061& 0.163& 1.52& 1.60\\[1ex]\hline
4 Fit & 29.7& 17.0& 12.2& 0.022& 0.068& -3.51& -2.99\\[1ex]\hline
5 Int. & 56.9& 21.7& 10.8& 0.036& 0.113& -0.87& 4.37\\[1ex]\hline
5 Fit & 17.9& 10.7& 7.60& 0.017& 0.046& -2.66& -2.53\\[1ex]\hline
6 Int. & 28.7& 17.5& 8.72& 0.029& 0.079& -2.64& 0.699\\[1ex]\hline
6 Fit & 10.2& 6.91& 5.88& 0.015& 0.033& -1.55& -0.287\\[1ex]\hline
\end{tabular}
\label{tab:runge-kxsq-asym}
\end{table}

For the RF analyses of interest in this work, global norms, such as the energy norm, converge well with very little difference between interpolated and fitted mesh geometry. However, it was found that for more localized quantities of engineering interest, the differences could be substantial. One such quantity of interest is the value of the integral of the electric field on critical surfaces, such as an RF antenna's Faraday grid.

\begin{figure}
\centering
\captionsetup{width=\linewidth, justification=centering}
	\includegraphics[width=0.98\textwidth]{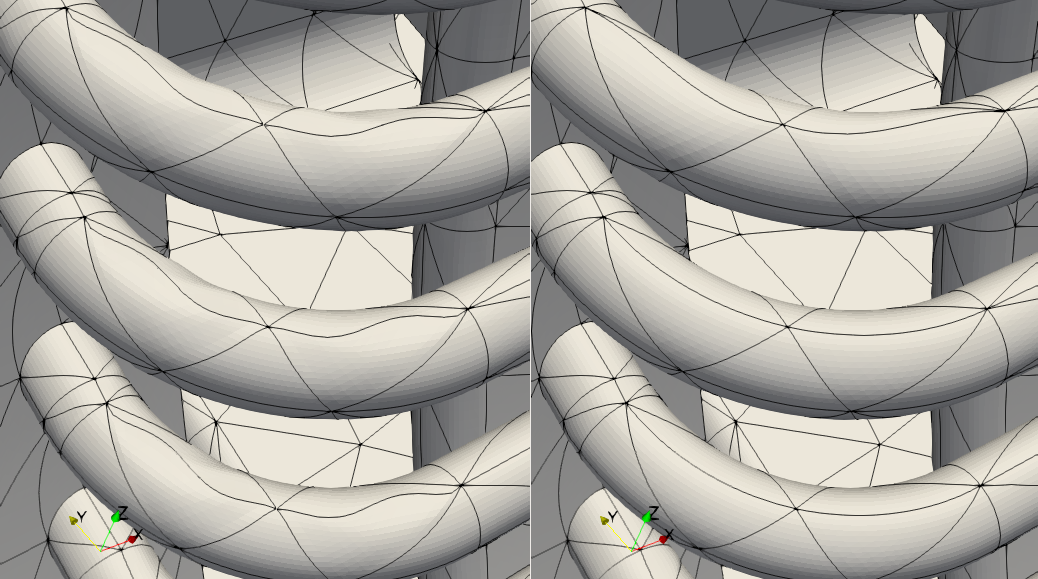}
	\caption{Close-up near CMOD RF antenna Faraday grid of the mesh geometry using fifth order B\'eziers and interpolation on left compared to curve fitting on right.}
	\label{fig:cmod-noise}
\end{figure}

\begin{figure}[htbp]
\centering
\captionsetup{width=\linewidth, justification=centering}
	\includegraphics[width=1.0\textwidth]{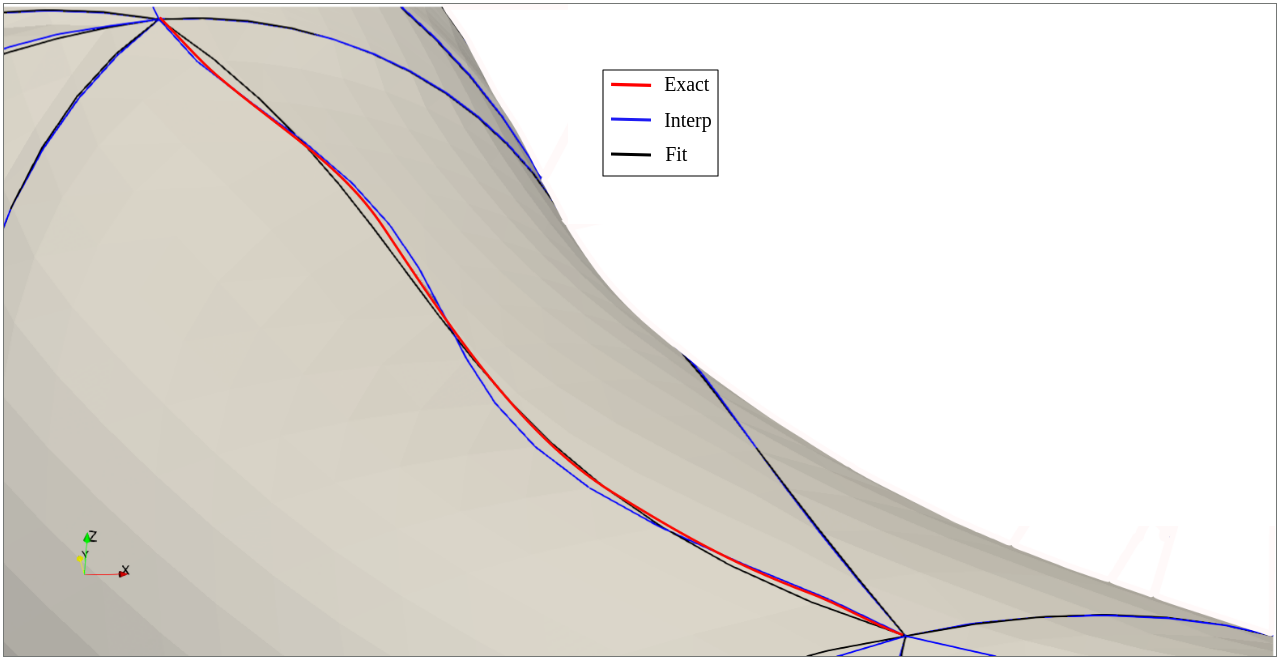}
	\caption{Close-up of elements on Faraday grid.}
	\label{fig:fgrid-overlap}
\end{figure}

Figure~\ref{fig:cmod-noise} shows a close-up of the surface mesh used for the simulations presented in~\cite{diab2026comparison}. The images on the left side of Figure~\ref{fig:cmod-noise} are based on interpolating mesh edges along the straight line in parametric space between the two mesh vertices, while the images on the right side use mesh edges fitted either to the model edge, or to the model face, on which the mesh edge is classified. It is clear from these images that the mesh edges on the toroidal portions of the Faraday grid are noisy. These are mesh edges classified on model faces where a straight line in parametric space is, in fact, a non-optimal curve fit for a mesh edge connecting the two mesh vertices.
This can be seen in Figure~\ref{fig:fgrid-overlap}, which shows the curve defined by following a straight line in parametric space, the fifth-order interpolation of the curve, and the curve defined by fitting a fifth-order curve to the surface. It was found that the noise introduced into the tangent vector directions, which increases with polynomial order, leads to overestimates of the integral of the electric field on the Faraday grid.
Table~\ref{tab:cmod-alpha} gives the maximum and mean differences in tangent angles for second through fifth-order element geometries for the mesh topology shown in Figure~\ref{fig:cmod-noise}. It can be observed that, as the element order increases, the tangent angle differences decrease for the fitted geometry. In contrast, for the interpolated geometry, the angle difference for the fourth-order element is worse than that of the second-order element, and the fifth-order element shows continued growth in the angle difference relative to the fourth-order element geometry.

\begin{table}
\centering
\captionsetup{width=\linewidth, justification=centering}
\caption{Angle $\alpha$ measured on Faraday grid of coarse mesh of CMOD RF antenna.}
\begin{tabular}{|l|c|c|}
\hline
 Order & Max $\alpha$ & Mean $\alpha$\\[1ex]
\hline\hline
 2 Int. & 28.1 & 13.1 \\[1ex]  \hline
 2 Fit & 26.4 & 12.4 \\[1ex]  \hline
 3 Int. & 24.2 & 11.3 \\[1ex]  \hline
 3 Fit & 21.6&  10.5\\[1ex]  \hline
 4 Int. & 41.3&  17.4\\[1ex]  \hline
 4 Fit & 18.4&  10.1\\[1ex]  \hline
 5 Int. & 44.8&  18.9\\[1ex]  \hline
 5 Fit & 16.2&  9.06\\[1ex]  \hline
 \end{tabular}
 \label{tab:cmod-alpha}
\end{table}

%% file: ShapeMeasure.tex
\section{Element Validity and Shape Measure}\label{ShapeMeasure}

\subsection{Element Validity}

The Jacobian matrix is defined as the first-order partial derivatives of the
physical coordinates with respect to the parametric coordinates (using indicial notation):
\begin{equation}
    J = \frac{\partial X_i}{\partial \xi_j}, \;\;\; i=x,y,z ;\;\;j=1,2,3;
\end{equation}
for $3D$ i.e., the dimension $d=3$, we can write the expanded form as:
\begin{equation}
    J = 
    \begin{bmatrix}
\frac{\partial X_x}{\partial \xi_1} & \frac{\partial X_x}{\partial \xi_2} & \frac{\partial X_x}{\partial \xi_3} \\
\frac{\partial X_y}{\partial \xi_1} & \frac{\partial X_y}{\partial \xi_2} & \frac{\partial X_y}{\partial \xi_3} \\
\frac{\partial X_z}{\partial \xi_1} & \frac{\partial X_z}{\partial \xi_2} & \frac{\partial X_z}{\partial \xi_3} 
\end{bmatrix}
\end{equation}
Since the derivative of a B\'ezier is a one order lower B\'ezier, the order of the terms in the Jacobian matrix are $\mathcal{O}(n-1)$ B\'eziers and determinant of $J$ is computed as follows.
\begin{equation}
\begin{aligned}
    det(J) = \frac{\partial X_x}{\partial \xi_1} \cdot \biggl(\frac{\partial X_y}{\partial \xi_2} \frac{\partial X_z}{\partial \xi_3} -  \frac{\partial X_y}{\partial \xi_3} \frac{\partial X_z}{\partial \xi_2} \biggl) -
                        \frac{\partial X_x}{\partial \xi_2} \cdot \biggl( \frac{\partial X_y}{\partial \xi_1} \frac{\partial X_z}{\partial \xi_3} - \frac{\partial X_y}{\partial \xi_3} \frac{\partial X_z}{\partial \xi_1} \biggl) \\
                      + \frac{\partial X_x}{\partial \xi_3} \cdot \biggl( \frac{\partial X_y}{\partial \xi_1} \frac{\partial X_z}{\partial \xi_2} - \frac{\partial X_y}{\partial \xi_2} \frac{\partial X_z}{\partial \xi_1} \biggl)
\end{aligned}
\end{equation}
$det(J)$ is the product of $d$ B\'ezier polynomials of order $n-1$, and is a $d*(n-1)$ order B\'ezier that is estimated by calculating corresponding control points $N_i^{(d(n-1))}$.
\begin{equation}
    det(J) = det(J)^{(d(n-1))}(\xi) = \sum_{i=1}^{q}b^{(d(n-1))}_{i}(\xi) \; N^{(d(n-1))}_{i}
\end{equation}
The details of the mathematical procedure and the formulation of the control points of $det(J)$ are discussed in references~\cite{George2012, george2016geometric, Johnen2013validty_remacle}. Since the $det(J)$ is expressed in B\'ezier form, its values are bounded by the maximum and minimum values evaluated at the control points of the order $d(n-1)$ B\'ezier polynomial. 
Consequently, a sufficient condition to ensure a positive Jacobian determinant for an order-$n$ curved simplex is that the minimum value of all control points of the determinant, $N_{min} = \min({N^{(d(n-1))}_{i}})$, is positive.
This method is conservative in that the actual value of $\min({det(J)})$ could be positive over the element domain while the minimum control point is negative. However, a negative minimum control point is an indication of a poor-quality element. This method is computationally efficient and effective for determining candidate mesh elements potentially associated with invalidity.

\subsection{Curved Element Shape Measures}
In addition to ensuring the validity of the elements in the mesh, it is important that all elements have a shape quality above some minimum threshold. Furthermore, it is desirable that the element shapes be as close to ideal as possible while accounting for the prescribed mesh size field. Unlike straight-sided elements, for which there exists a set of equivalent element shape quality measures, the definition of shape measures for curved elements has proven more challenging. What is common among the various measures used~\cite{loseille_measure, feuillet2020optimization, roca_distortion, Johnen2013validty_remacle, lu2014parallel, stees_distortion} is the use of multiple terms that consider both the underlying anisotropy of the element relative to the requested level of anisotropy and a measure of distortion caused by curving the element.

One straightforward measure of the distortion of a curved element is the ratio of the minimum and maximum determinants of the element, $Q_c$:
\begin{equation}
Q_c=\left(\frac{\min_{\xi \in \Omega^{e}}{det(J(\xi))}}{\max_{\xi \in \Omega^{e}}
{det(J(\xi))}}\right)^{1/d}
\end{equation}

An obvious problem with this measure is the complexity and cost of determining $\max_{\xi \in \Omega^{e}}$ and $\min_{\xi \in \Omega^{e}}$. 
Since $det(J)$ is bounded by the maximum and minimum values evaluated at the control points of the order $d(n-1)$ B\'ezier polynomial representing $det(J)$, the less expensive quantity to compute, $\frac{N_{min}}{N_{max}}$, can serve as the distortion measure, where $N_{min}$ and $N_{max}$ are the smallest and largest values of the control coefficients of the Jacobian determinant. These values bound $det(J)$~\cite{Farin1992, loseille_measure, george2016geometric, Johnen2013validty_remacle, lu2014parallel}.
\begin{equation}
    N_{min} \leq det(J)\leq N_{max}
\end{equation}
The $d^{th}$ ($d =$ spatial dimension) root of the distortion normalizes the measure to account for the spatial dimension of the curved mesh entity~\cite{loseille_measure}.
\begin{equation}
\label{eq:Qc}
    Q_c = \Biggl(\frac{N_{min}}{N_{max}}\Biggl)^{1/d}
\end{equation}
The range $Q_c$ is within $[0, 1]$ for elements that are considered valid curved elements, and it is negative for what are considered as invalid elements.

While the $Q_c$ measure considers the shape deviation of a curved element with respect to its underlying straight-sided counterpart, it does not account for the shape quality of the underlying straight-sided element itself, as noted in reference~\cite{loseille_measure}. Consequently, the metric $Q_c$ reports the optimal value of 1 even when the straight-sided tetrahedron is close to degenerate.
Therefore, the shape measure employed here, $Q$, is defined as the product of $Q_c$ and $Q_s$, where $Q_s$ is one of the standard straight-sided element shape measures.
\begin{equation} \label{eq:qsqc}
     Q=Q_s*Q_c
\end{equation}
The specific straight-sided element measure used is the mean ratio~\cite{compere2010mesh, ibanez2016conformal, Li2005, Liu1994a} which for tetrahedrons is:
\begin{equation} \label{eq:tet_mean_ratio}
Q_s = \left(\frac{V_s}{\gamma_s\cdot l_{s-\text{RMS}}^3}\right)^{\frac23},\,
l_{s-\text{RMS}}=\left(\frac16\sum_{i=1}^6 \left(l_{s,i}\right)^2\right)^\frac12, \,
\gamma_s = \frac{1}{\sqrt{72}}
\end{equation}

In equation~\ref{eq:tet_mean_ratio}, $V_s$ is the straight-sided tetrahedron volume, $l_{s,i}$ is the straight-sided length of edge $i$ of the tetrahedron, and $\gamma_s$ is the volume of an equilateral tetrahedron with unit edge length.
An equilateral tetrahedron has $Q=1$, while a tetrahedron with a $Q \le 0$ is considered an invalid element. Acceptable tetrahedron have $Q \ge Q_{min} > 0$.

%% file: CurvedMeshAdaptOverview.tex
\section{Curved Mesh Adaptation Overview} \label{AdaptOverview}
As overviewed in section \ref{Introduction} there are alternative strategies that have been taken to the development of a curved mesh adaptation procedure, each having relative advantages depending on the requirements of the target applications. The curved mesh adaptation procedures presented in this paper are targeted for applications for which:

\begin{itemize}
    \item The accurate transfer of solution fields to the adapted mesh, while accounting for conservation constraints, can be effectively executed. Cavity-based mesh adaptation supports local solution transfer operations that have been demonstrated to be more efficient and accurate than global mesh-to-mesh transfer.
    
    \item The accuracy of local quantities of interest can be adversely influenced by curved element geometries that do not maintain sufficient geometric approximation accuracy and/or are not sufficiently smooth (see Section~\ref{CompareFit}).
    
    \item Elements in the adapted mesh may be large with respect to the local geometry, forcing the need to curve some of the interior mesh edges and faces to ensure element validity and acceptable element shapes.
    
    \item The complexity of the domain and physics being solved is such that meshes containing millions of elements are often required. For the efficiency of both the simulation and mesh adaptation procedures, it is desirable to employ straight-sided planar-faced elements so long as they have good shape quality and do not introduce significant geometric approximation errors.
\end{itemize}

Based on these requirements, the key features of the present curved mesh adaptation procedure are:
\begin{itemize}
    \item Cavity-based mesh modifications that can be effectively executed using two levels of parallelism on heterogeneous supercomputers~\cite{ibanez2016conformal}.
    
    \item The mesh edges and faces classified on the model boundary are curved using the geometric approximation procedure presented in Section~\ref{sec:crv-fit} to the order of the finite elements used. Currently, support is provided for orders up to six, since this is the highest order employed in the current target applications.
    
    \item The order of interior mesh entities is limited to three for computational efficiency, and interior mesh entities are maintained at order one whenever acceptable element shape quality can be preserved.
\end{itemize}

Algorithm~\ref{alg:crvAdaptOverall} indicates the main steps in an automated adaptive simulation. Given the problem definition in terms of an attributed CAD model, the first step is the generation of an initial mesh (Line 3). Since most available automatic mesh generators are limited to linear or quadratic element shapes, the mesh edges classified on curved model boundaries need to improve their level of geometric approximation to match that of the finite element(s) they bound (Line 4). 
In general, the process of changing the shape of mesh edges and faces can degrade the shape quality of the elements they bound or even render them invalid. To address this issue, the elements connected to mesh edges and faces whose shapes have changed are checked to determine whether their shapes are acceptable. The \textbf{Fix Element Shape} procedure is invoked for any unsatisfactory elements (Line 5).

Given an acceptable initial mesh, the analysis is started. To support cases with multiple analysis steps, a while loop (Lines 6-16) is introduced. After each analysis step is executed, a posteriori error estimates are evaluated and used to calculate a new mesh metric field (Line 8).
Lines 9-14 define the while loop that adapts the mesh through the application of mesh modification operations. In each pass through that loop, one level of \textbf{Element Collapse} and \textbf{Element Refinement}, followed by a loop of \textbf{Fix Element Shape}, is carried out.

\begin{algorithm}
	\caption{Overall adaptive simulation workflow}
	\begin{algorithmic}[1]
	    \State INPUT: A simulation problem defined in terms of a CAD model attributed with an initial set of mesh control parameters and the attributes (material properties, loads and boundary conditions) required for the analysis to be carried out.
	    \State OUTPUT: Simulation results for the requested quantities of interest. \bigskip
        \State Invoke an automatic mesh generator to generate the initial mesh.
        \State Inflate the geometric order of mesh edges and faces classified on curved model edges or faces ($M_i^1 \sqsubset G_j^l$, $l=1 \text{ or }2$ and $M_i^2 \sqsubset G_j^2$) to the order of the finite elements they bound.
        \State Check the validity and shape quality of elements connected to mesh edges and faces that have had their geometry inflated and perform shape improvement of any unsatisfactory elements. 
        \While {Additional analysis steps are required}
            \State Execute the next analysis step, or indicated set of steps. 
            \State Perform a posteriori error estimation and calculate new target mesh  metric field.
            \While {Target metric not satisfied}
                \State Perform a loop of \textbf{Coarsen} performing mesh entity collapse operations to coarsen the mesh.
                \State Perform loop of \textbf{Refinement} using entity split operations to make the  mesh finer.
                \While{Elements with unacceptable element quality exist}
                    \State Execute \textbf{Fix Element Shape} to ensure the shape quality 
                    of newly created elements.
                \EndWhile
            \EndWhile
        \EndWhile
    \end{algorithmic}
    \label{alg:crvAdaptOverall}
\end{algorithm}

%% file: CoreMeshAdaptOperators.tex
\section{Curved Mesh Entity Operations}\label{CoreOperators}

The curved mesh adaptation procedure builds on a set of procedures involved in the definition of curved mesh entities. These procedures include:
\begin{itemize}
\item Curving the mesh entities classified on curved model boundaries to the required level of geometric approximation while ensuring that the mesh entities remain valid and maintain acceptable shape quality.
\item Defining the shape of new mesh entities when performing entity split operations.
\item Defining the shape of newly created curved mesh entities within a cavity while maintaining acceptable element shapes.
\end{itemize}

\subsection{Curving Mesh Entities on Curved Model Boundaries} \label{CurvingToBoundary}

The procedure used to take an initial input mesh and curve it to the desired order on the boundary is shown in Algorithm \ref{alg:inflOrder}.
In Lines 3-9, a loop over the model edges is executed. If a model edge is curved, reverse classification is employed (loop in Lines 5-7) to curve the mesh edges classified on the curved model edges by invoking the least-squares fitting procedure, as stated in Line 6.
In Lines 10-19, a loop over the model faces is executed. If a model face is curved, the mesh edges classified on those model faces are first curved in Lines 12-14, and then the mesh faces classified on those model faces are curved, as stated in Lines 15-17.

Algorithm \ref{alg:leastSquares} provides an overview of the fitting procedure. First, $m$ points used to fit the mesh entity $M_i^{d_i}$ are constructed by querying the geometric model entity $G_j^{d_j}$ in Lines 3-4. The matrices for the Bernstein polynomials, $B$, and the control points, $C$, are formulated in Lines 5-6. In Lines 7-8, the internal control points are computed by solving the linear system and stored with $M_i^{d_i}$.

\begin{algorithm}
  \caption{Construction of mesh geometry for mesh entities classified on boundary}
  \begin{algorithmic}[1]
      \State INPUT: Input initial mesh with the available geometric shape information.
      \State OUTPUT: Curved mesh entities on boundary constructed to desired order.\bigskip

        \For{$G_j^1 \gets \{G^1\}$} \hfill $\triangleright$ loop over model edges
        \If{$G_j^{1}$ is curved}
            \For{$M_i^1 \gets \{M^{1}\} \sqsupset G_j^1$} \hfill $\triangleright$ loop over mesh edges classified on model edge
                \State Curve $M_i^1$ using Algorithm \ref{alg:leastSquares}.
            \EndFor
        \EndIf
        \EndFor

        \For{$G_j^2 \gets \{G^2\}$} \hfill $\triangleright$ loop over model faces
        \If{$G_j^{2}$ is curved}
            \For{$M_i^1 \gets \{M^{1}\} \sqsupset G_j^2$} \hfill $\triangleright$ loop over mesh edges classified on model face
                \State Curve $M_i^1$ using Algorithm \ref{alg:leastSquares}.
            \EndFor

            \For{$M_i^2 \gets \{M^{2}\} \sqsupset G_j^2$} \hfill $\triangleright$ loop over mesh faces classified on model face
                \State Curve $M_i^2$ using Algorithm \ref{alg:leastSquares}.
            \EndFor
        \EndIf
        \EndFor
       \end{algorithmic}
    \label{alg:inflOrder}
\end{algorithm}

\begin{algorithm}
  \caption{Procedure to fit mesh entity geometry to model boundary}
  \begin{algorithmic}[1]
      \State INPUT: Mesh entity $M_i^{d_i}$ classified on model entity $G_j^{d_j}$.
      \State OUTPUT: Mesh entity geometry constructed to desired order $n$.\bigskip

            \State Construct $m$ points $\bm{\xi}$ in parametric space of model entity $G_j^{d_j}$.
            \State Use $\bm{\xi}$ to query geometric model entity $G_j^{d_j}$ for points $\bm{D}$ in real space.
            \State Evaluate Bernstein polynomials ${b}^{(n)}$ and assemble Bernstein matrix $\bm{B}$.
            \State The linearized system for control points $\bm{C}$ is formulated.
            \State Compute internal control points by solving $\bm{B}^T\bm{B}\;\bm{C} = \bm{B}^T\bm{D}$.
            \State Store control points and geometric order $n$ with $M_i^{d_i}$.
        
       \end{algorithmic}
    \label{alg:leastSquares}
\end{algorithm}

\subsection{Procedure to Subdivide Curved Elements in Cavity} \label{sec:subdivide-curve}

When mesh entities are subdivided during a refinement operation, the first step in the subdivision process performs entity splitting in the parametric space of the elements being subdivided. The key reason for performing the initial subdivision operation in this manner is that the resulting elements are guaranteed to be valid. In cases where an already curved element is subdivided, the initial versions of the new mesh entities will also be curved.
Given this set of valid mesh entities, the next step evaluates the resulting element quality when the newly created mesh edges are made straight and the mesh faces are made planar in physical space. If the element quality is satisfactory, the edges remain straight and the faces planar; otherwise, they are left curved.

\begin{figure}
\centering
\captionsetup{width=\linewidth, justification=centering}
    \centering
	\setlength{\fboxrule}{0.5pt}\fbox{\includegraphics[width=0.75\textwidth]{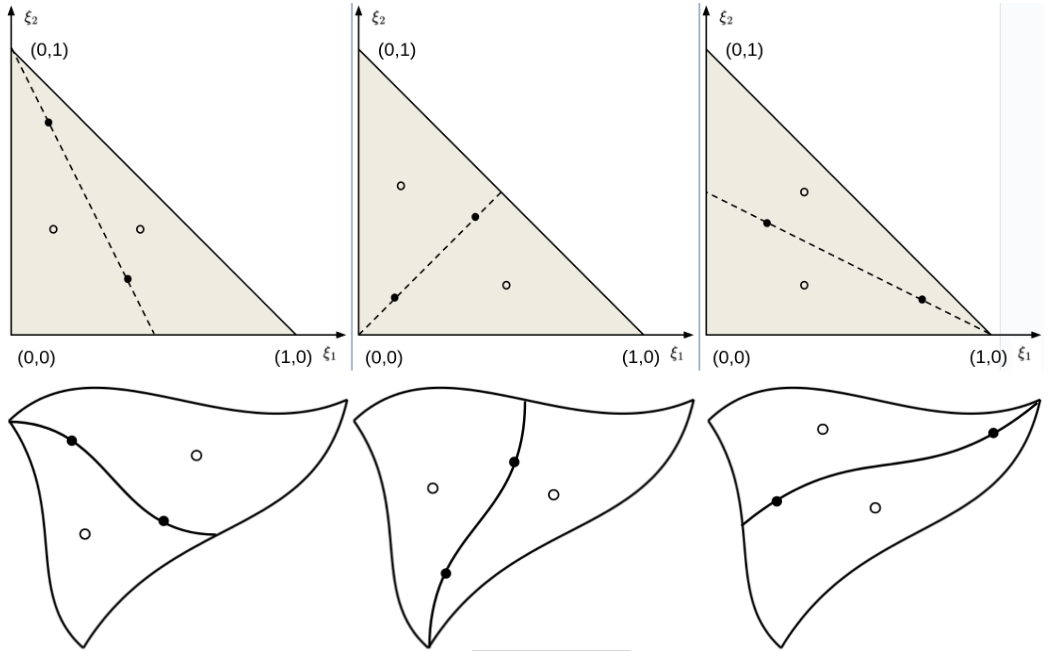}}
	\caption{High order points of new entities created after subdivision will have different coordinates in reference to the proper orientation of the old triangle with respect to parametric space.}
	\label{fig:tri_split_param}
\end{figure}

In the first step of the subdivision process, the parametric information for the new entities is templated based on the relative orientation of the new entities with respect to the reference entity (edge, triangle, or tetrahedron) that exists before the split operation.
Figure~\ref{fig:tri_split_param}  illustrates this process for a cubic triangle. 
After determining the parametric locations, $\bm{\xi}_{old2new}$, for the new points in the original triangle, the physical coordinates are calculated by querying the B\'ezier form of the original triangle.

\begin{equation}
\textbf{X}^{n}(\bm{\xi}_{old2new}) = \sum_{i=1}^q\textbf{C}^{n}_{i}b^{n}_{i}(\bm{\xi}_{old2new})
\end{equation}

\begin{figure}
\centering
\captionsetup{width=\linewidth, justification=centering}
	\setlength{\fboxrule}{0.5pt}\fbox{\includegraphics[width=.3\textwidth]{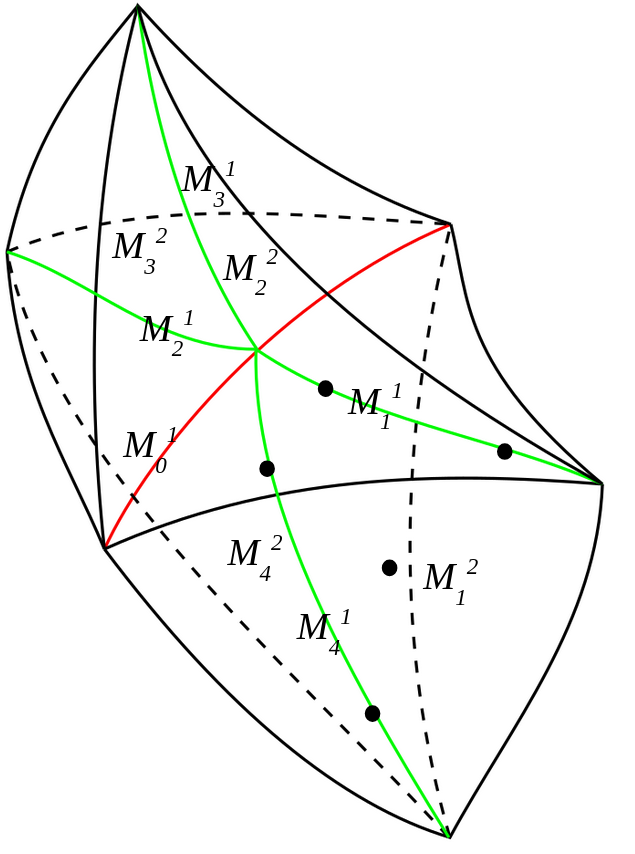}}
	\caption{Illustration of creating new entities in cavity using subdivision by edge split (red) in $3D$.}
	\label{fig:tet_split}
\end{figure}

Figure~\ref{fig:tet_split} shows an edge split for a $3D$ curved tetrahedral mesh cavity comprised of tetrahedra adjacent to the edge being split.
The edge $M_0^1$ being split is marked in red, and the newly created edges $M_1^1$, $M_2^1$, $M_3^1$, and $M_4^1$ are marked in green.
The high-order points for the new face $M_1^2$ are shown along with those of the adjacent edges $M_1^1$ and $M_4^1$. The mesh geometry information for the new edges can be obtained by querying the faces adjacent to the edge being split, and the shape information for the new faces can be obtained from the adjacent tetrahedra in the original cavity.
In the second step of the process, the element qualities are checked for the case that the newly created mesh edges and faces are made straight and planar. If the elements shapes are acceptable, the straight edges and planar faces are retained.

\subsection{Interior Mesh Entity Curving Procedure} \label{InteriorCurving}

One approach to curving interior mesh entities is to employ numerical optimization procedures \cite{stees_distortion, persson2024local, loseille_opt_cavity2023, dobrev2022adapt-optim}. Even when applied iteratively at the cavity level, these methods tend to have a high computational cost that grows rapidly as the order of the mesh entity geometry increases.
An alternative approach is taken in the current work. In this approach interior mesh entity curving is limited to improving the shape of only those elements with unacceptable shape metrics, and the order of curved mesh entities is limited to third order (cubic).
In addition to avoiding the added cost associated with computing the larger number of shape parameters required for entities of order higher than three, the four shape parameters associated with a cubic B\'ezier mesh edge can be easily understood and controlled while still providing a substantial degree of shape flexibility due to the ability of the curve to exhibit changes in curvature. The four parameters defining the shape of a cubic edge are the directions of the tangents of the cubic at the edge's vertices and the locations of the B\'ezier cubic control points along those tangents.

\begin{figure}
\centering
\captionsetup{width=\linewidth, justification=centering}
\setlength{\fboxrule}{0.5pt}\fbox{\includegraphics[trim={19.5cm 1.4cm 0 0.15cm}, clip, width=0.3\textwidth, angle=0]{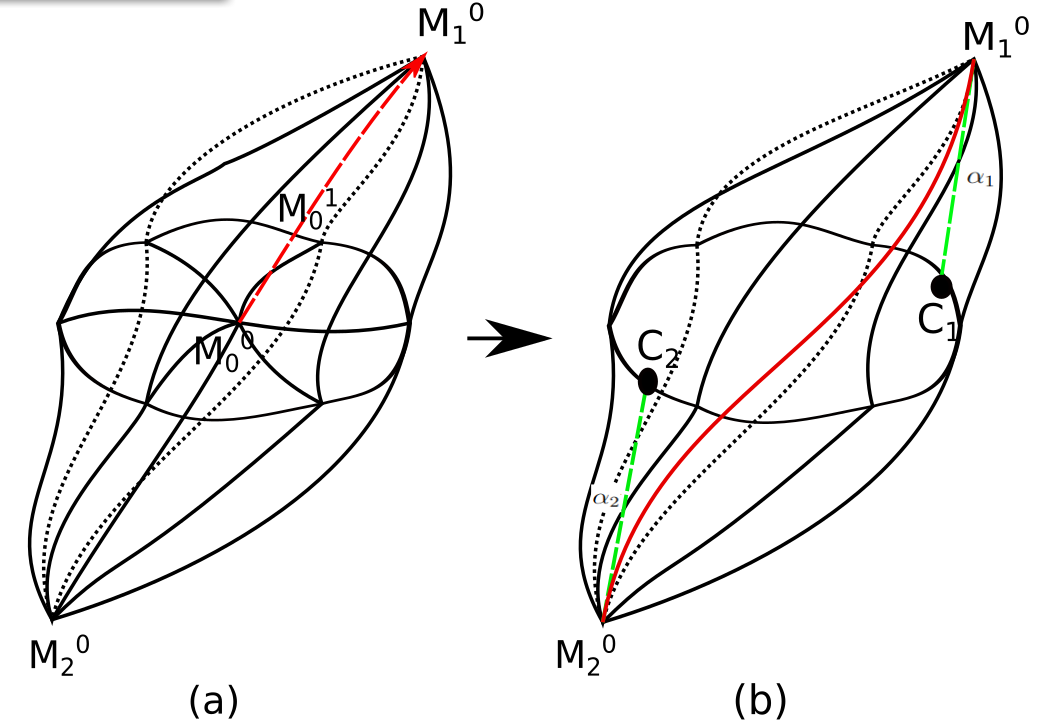}}
\caption{Using information from cavity boundary to place control points on the average tangents of the upper and lower faceted cones.} \label{fig:tetlower}
\end{figure}

Referring to the mesh cavity shown in Figure~\ref{fig:tetlower}, the cubic edge being defined connects $M_1^0$ and $M_2^0$. The tangent directions at $M_1^0$ and $M_2^0$ are defined as the averages of the tangents of the cavity mesh edges incident to $M_1^0$ and $M_2^0$, respectively. The distances along the tangents to $M_1^0$ and $M_2^0$, denoted by $\alpha_1$ and $\alpha_2$, determine the locations of the control points $C_1$ and $C_2$.
The locations of the control points dictate the relative sharpness of the cubic $S$-curve. The goal is to set the values of $\alpha_1$ and $\alpha_2$ such that the shape measure of the worst-shaped element in the cavity is maximized. Since this is only a two-parameter optimization problem, optimization methods or relatively simple heuristics can be applied to determine $\alpha_1$ and $\alpha_2$.
The cubic faces bounded by the cubic edge are created using a blending function~\cite{Dey1997a} to define the interior control point.

\section{Mesh Modification Operators}\label{ModificationOperators}

The cavity-level transformation from the input mesh to the output mesh can be expressed as a series of local cavity modifications, each of which consists of the removal of a small number of local mesh entities followed by the addition of a small number of new mesh entities that fill the cavity.

The cavity-based mesh modification operators include:
\begin{enumerate}
\item Refinement: Create a finer mesh by splitting the indicated mesh edges.
\item Coarsening: Create a coarser mesh by collapsing the indicated mesh edges.
\item Swapping: Modify the mesh topology of the cavity to better match the mesh size field or improve the desired quality metric.
\end{enumerate}

The following steps are executed as part of the generalized curved mesh modification process:
\begin{enumerate}
\item Test whether the desired local mesh topology modification is applicable by applying topological checks \cite{Li2005}.
\item If no topological violation is found, apply the appropriate local mesh modification operations and perform curving of entities on the boundary, as discussed in Section \ref{sec:crv-fit}.
\item Check the element quality, considering the curved geometry on the boundary and treating the new interior entities as straight-sided.
\item If an inverted or poor-quality element is detected, apply the mesh curving algorithms discussed in Section \ref{InteriorCurving}. Further quality improvement can be achieved through the application of \textbf{Fix Element Shape}.
\end{enumerate}

In some cases, a combination of operators can be used to eliminate sliver elements or to create space for the generation of valid elements during mesh curving procedures. The following sections describe the procedures for the core mesh modification operators.

\subsection{Refinement}
\begin{figure}
\centering
	\includegraphics[width=0.65\textwidth]{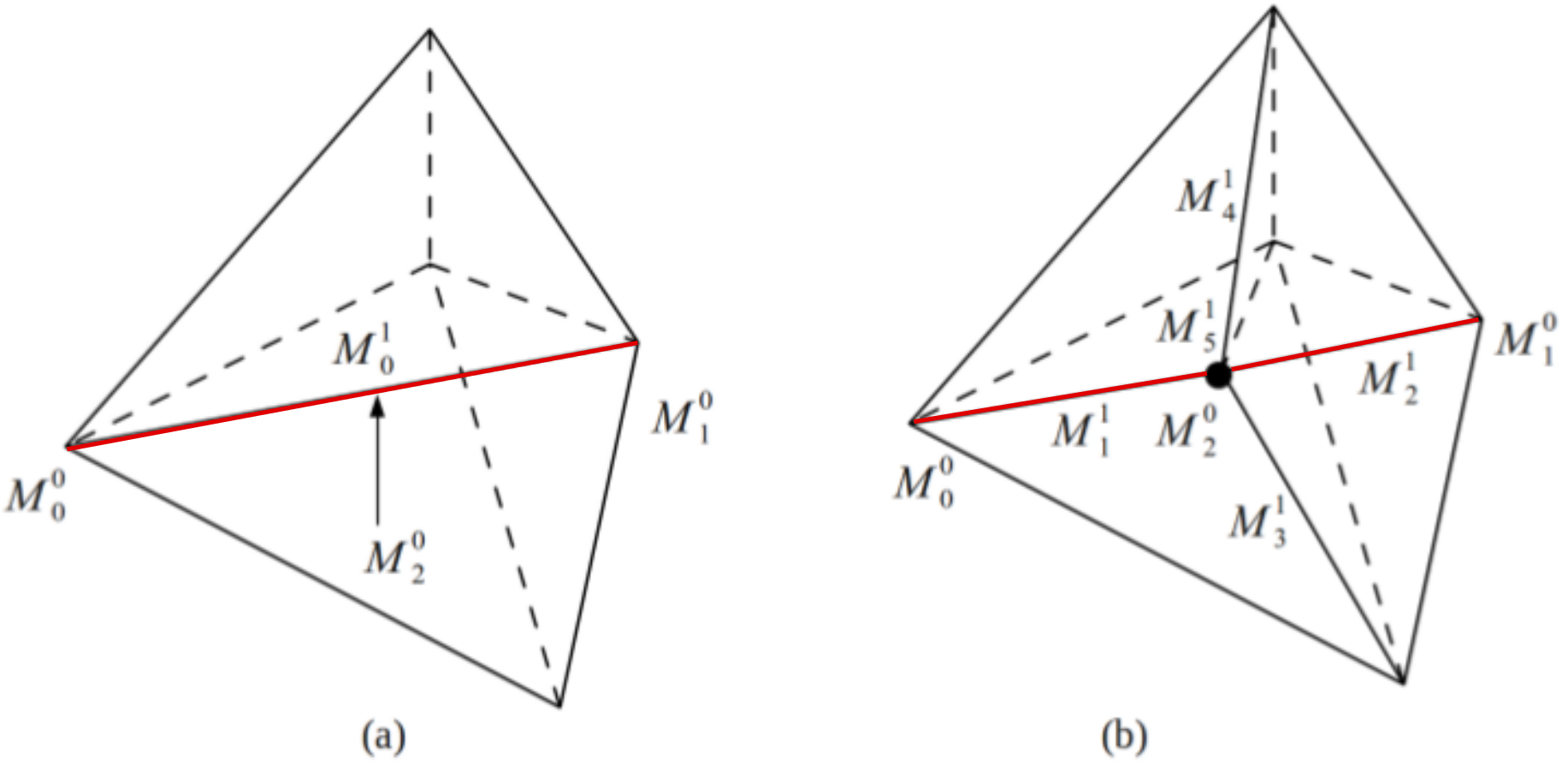}
	\caption{Edge split operation: (a) $M_0^1$ is targeted to be split by introducing a new vertex $M_2^0$. (b) New entities created after edge split.}
	\label{fig:split-topo}
\end{figure}

Mesh refinement is implemented using edge splits. Figure~\ref{fig:split-topo} illustrates the topological changes caused by an edge split applied to two adjacent elements. The selected edge is bisected, and the adjacent elements are subdivided by inserting a new vertex that splits the edge and the higher-dimensional entities it bounds.
Algorithm~\ref{alg:refinement} provides an overview of the mesh refinement procedure. During the execution of a refinement step, all mesh edges whose lengths exceed a threshold value greater than $\sqrt{2}$ in metric space are selected for splitting, as indicated in Line 3.
An independent set of non-overlapping cavities is then determined~\cite{Cougny1999} in Line 5 so that the refinement operations may be applied in parallel using multithreaded execution~\cite{ibanez2016conformal}.

\begin{algorithm}
  \caption{Refinement of a curved mesh based on metric field}
  \begin{algorithmic}[1]
      \State INPUT: Mesh and metric field attributed with target edge lengths.
      \State OUTPUT: Refined mesh.\bigskip

        \State Compare current edge lengths with target lengths and mark long edges.
        \While{There are long edges to be split}
        \State Define independent sets of cavities including consideration of mesh partition boundaries.
        \For{Loop over cavities}
            \State Create topology and classification for new entities by splitting old edge in the cavity.
            \State Maintain geometric order for any required new curved $M_i^1$ and $M_i^2$ using respective old entity through subdivision (see Section \ref{sec:subdivide-curve}).

            \For{$M_i^0 \gets M_i^0 \sqsubset G_j^l \;(l=1 \text{ or }2)$} \hfill $\triangleright$ loop over new vertices classified on model boundary
                \State Place $M_i^0$ on model boundary by making parametric query.
            \EndFor
            \For{$M_i^1 \gets M_i^1 \sqsubset G_j^l \;(l=1 \text{ or }2)$} \hfill $\triangleright$ loop over new edges classified on model boundary
                \State Curve $M_i^1$ to model boundary (see Algorithm \ref{alg:leastSquares})
            \EndFor
            \For{$M_i^2 \gets M_i^2 \sqsubset G_j^2$} \hfill $\triangleright$ loop over new faces classified on model boundary
                \State Curve $M_i^2$ to model boundary (see Algorithm \ref{alg:leastSquares})
            \EndFor

            \State Check quality and if new curved $M_i^1$ are made straight-sided and new curved $M_i^2$ made planer yields satisfactory shapes execute that.
            
            \For{Undesirable element shapes}
                \For{$M_i^l \gets M_i^l \sqsubset G_j^3 \; (l=1 \text{ or }2)$} \hfill $\triangleright$ loop over new interior edges and faces
                    \State Curve $M_i^l$ in the cavity (see Section \ref{InteriorCurving}).
                \EndFor
            \EndFor
        \EndFor
        \EndWhile

       \end{algorithmic}
    \label{alg:refinement}
\end{algorithm}
The new vertex inherits the classification of the edge being split, as indicated in Line 7. In Line 8, new curved entities are created by subdividing the mesh entities bounded by the edge using the procedure discussed in Section~\ref{sec:subdivide-curve}, and new entities on boundary are curved as stated in Lines 9-17.
In Line 18, a procedure is executed to check whether making the new interior curved edges straight and the new curved faces planar results in elements of satisfactory quality. If this produces acceptable element shapes, then those edges are kept straight and the faces are kept planar.
After inflation to the boundary, if undesirable element shapes remain in the cavity, the new interior edges and faces, which are initially straight-sided, are curved as discussed in Section~\ref{InteriorCurving}, and stated in Lines 19-23.

\begin{figure}
\centering
\begin{subfigure}{.49\textwidth}
  \centering
  \includegraphics[width=.9\textwidth,  trim={0cm 0cm 5cm 0cm}]{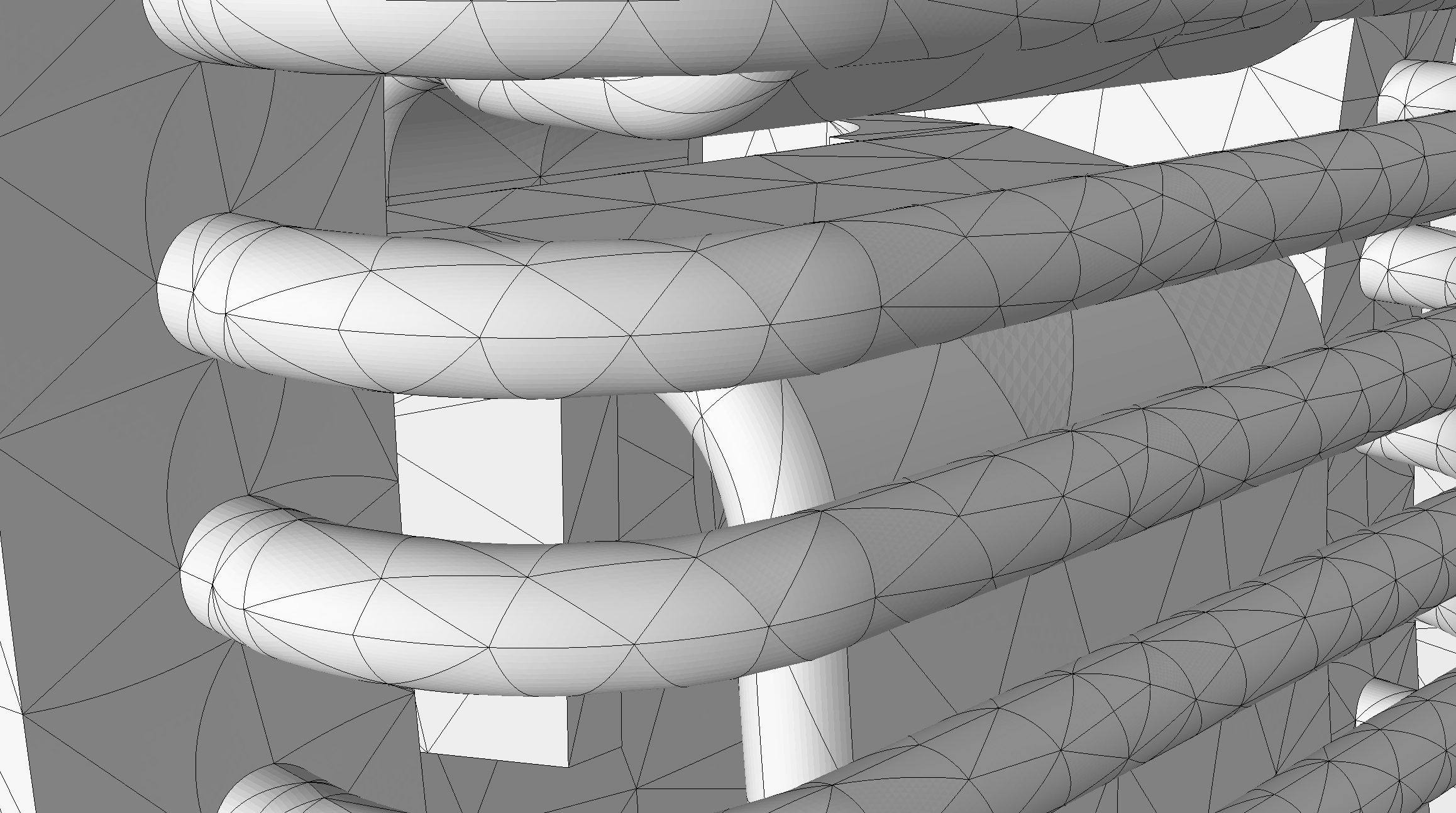}
  \caption{}
\end{subfigure}
\begin{subfigure}{.49\textwidth}
  \centering
  \includegraphics[width=.9\textwidth,  trim={0cm 0cm 5cm 0cm}]{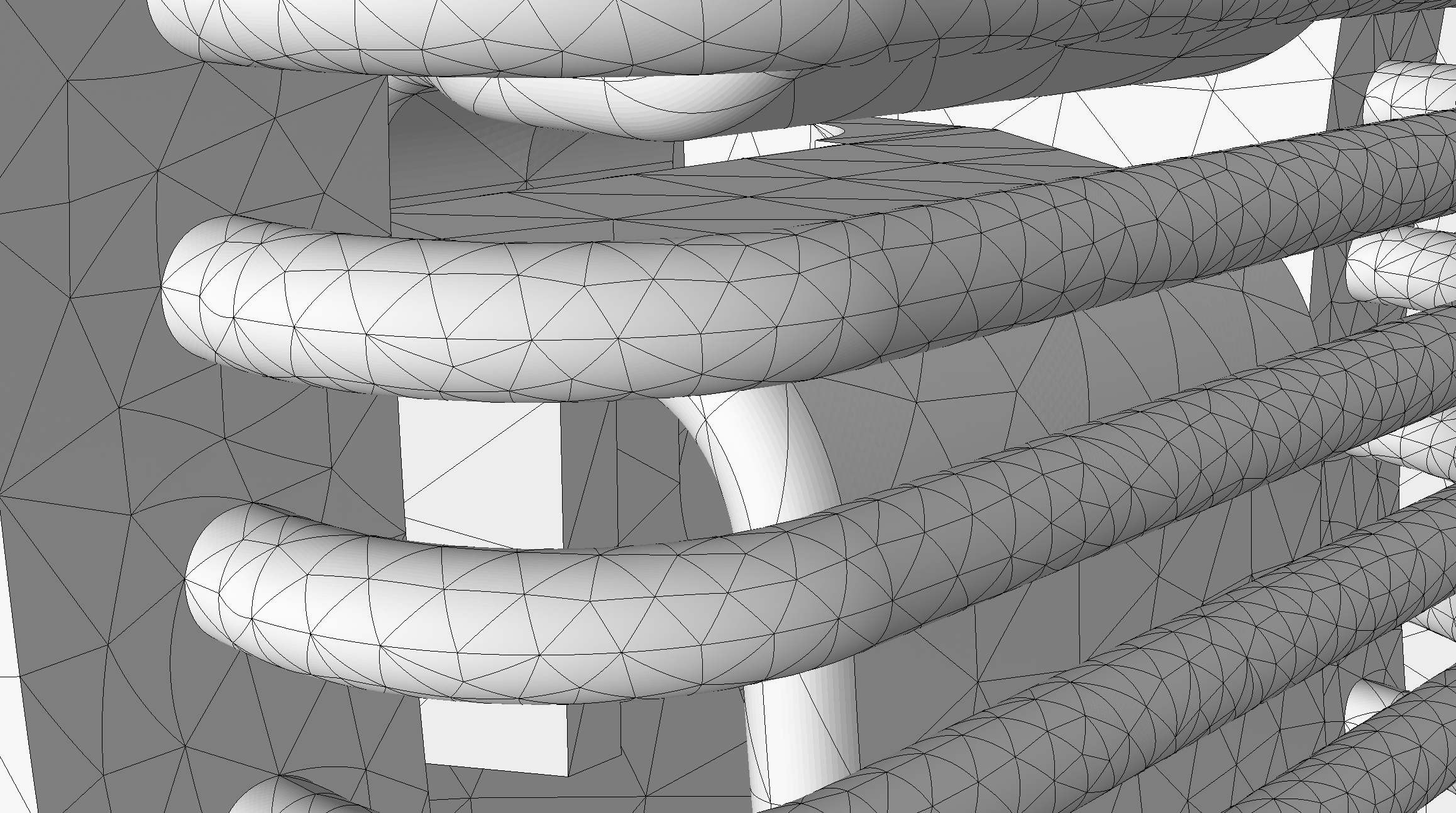}
  \caption{}
\end{subfigure}
\caption{Demonstration of refinement near the Faraday grid (cylindrical tubular structures): (a) initial and (b) refined mesh.}
\label{fig:refine-fscr}
\end{figure}

Figure~\ref{fig:refine-fscr} shows a close-up view of a coarse cubic mesh geometry before and after refinement on the Faraday screen of the C-Mod RF antenna.

\subsection{Coarsening}

\begin{figure}
\centering
	\includegraphics[width=0.65\textwidth]{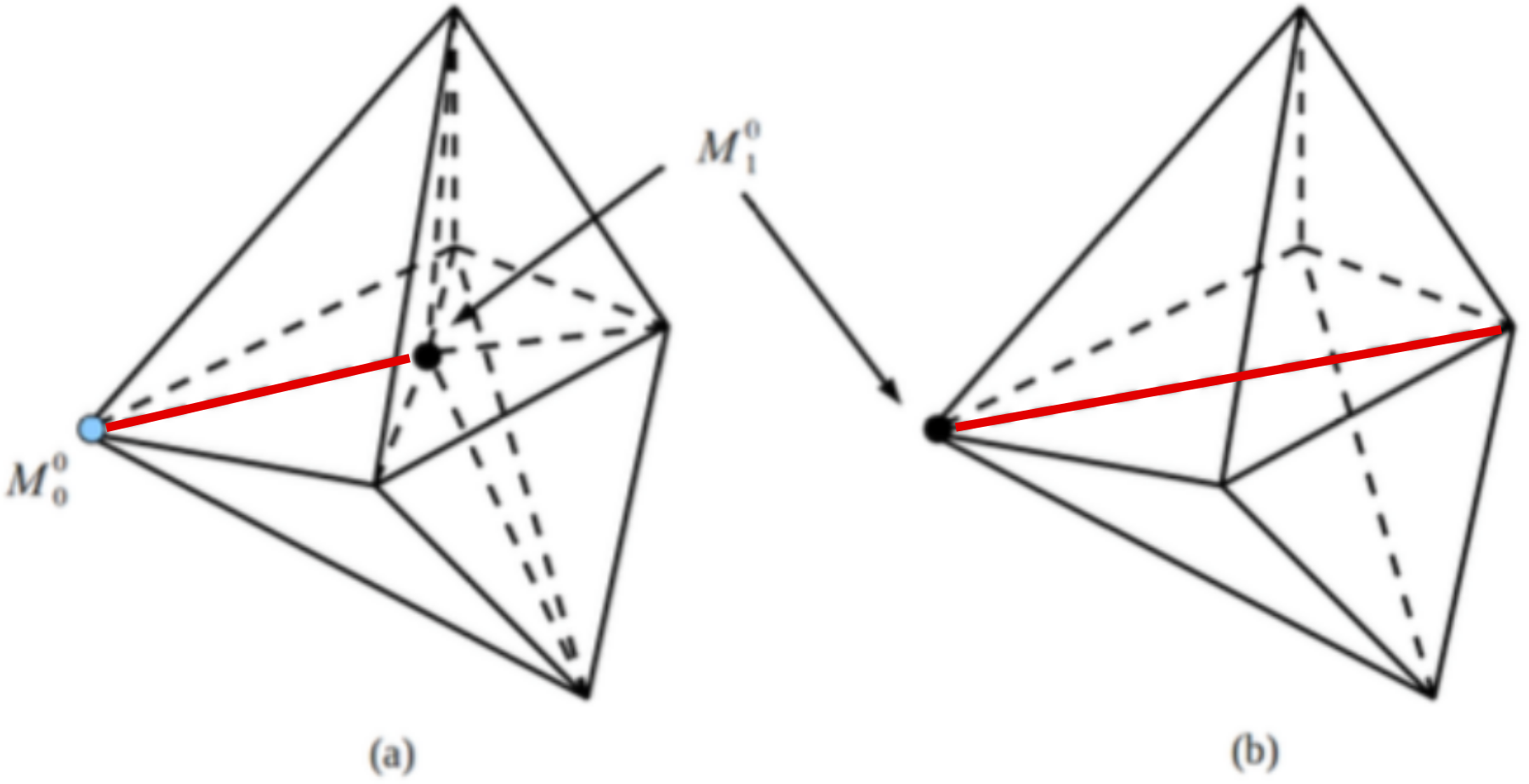}
	\caption{Edge collapse in a simple cavity: (a) The edge between $M_0^0$ and $M_1^0$ is the edge to be collapsed. (b) Edge is collapsed by moving $M_1^0$ to $M_0^0$.}
	\label{fig:collapse-topo}
\end{figure}

Coarsening is implemented by collapsing edges that are shorter than $1/\sqrt{2}$ in metric space.
In an edge collapse operation, a mesh vertex is moved onto the mesh vertex at the other end of the mesh edge, thereby collapsing that edge and its adjacent faces and regions. 
An example of the changes to mesh topology and geometry during an edge collapse for a $3D$ tetrahedral mesh is shown in Figure~\ref{fig:collapse-topo}, where the edge is collapsed by moving vertex $M_1^0$ to vertex $M_0^0$.

An overview of the coarsening procedure for curved meshes is shown in Algorithm~\ref{alg:coarsening}.
All edges whose metric lengths are below the specified threshold, $1/\sqrt{2}$, are marked as ``short'' edges that are candidates for collapse, as stated in Line 3. A short edge can be collapsed from either vertex. Thus, the focus is shifted from the short edges to their vertices such that each collapsing vertex identifies and represents an adjacent collapsing edge.

If a vertex has multiple adjacent short edges that may be collapsed, the first criterion is to select the shortest adjacent edge, as stated in Line 4. If the edge lengths are equal within a tolerance of $10^{-15}$, preference is given to the collapse that produces better-shaped elements in the cavity while keeping the new entities straight. The collapses in the allowed directions are evaluated using validity and quality criteria for curved elements. All other possible collapses involving the vertex as an endpoint are discarded.

Therefore, for edge collapses, the cavity is defined by the combination of an edge and one of its endpoint vertices. The independent set selection employed in Line 6 operates on maps of vertices as the keys and their selected post-collapse qualities as the values. The selected independent set of vertices is then collapsed along each vertex's chosen edge. During the collapse operation, mesh classification is maintained. The preservation of topological similarity to the CAD model in Lines 8-9 is ensured using the checks described in reference~\cite{Li2005}.
These checks prevent collapses from the boundary into the interior, as well as collapses that could create zero-volume elements.

The new edges and faces classified on the boundary are curved to conform to the model geometry using the procedure discussed in Section~\ref{sec:crv-fit}, as stated in Line 10. If undesirable element shapes exist in the cavity, the new interior edges and faces are curved using the procedure discussed in Section~\ref{InteriorCurving}.
The ability to curve the new interior edges and faces provides an efficient means of improving element quality during coarsening.
\begin{algorithm}
  \caption{Coarsening of a curved mesh}
  \begin{algorithmic}[1]
      \State INPUT: Mesh and metric field attributed with target edge lengths.
      \State OUTPUT: Coarser mesh than the input.\bigskip

        \State Compare current edge lengths with target lengths and mark short edges and their adjacent vertices.
        \While{There are short edges to be collapsed}
        \State Prioritize vertex collapse along shortest adjacent edge length and then best resultant quality.
        \State Form independent sets of cavities including consideration of mesh partition boundaries.
 
        \For{Loop over cavities}
            \State Create topology and classification for new entities using the old cavity and initialize new entities straight-sided.
            \State If collapse from boundary to interior and/or zero volume, filter out that short edge and move to next cavity.
            \State Execute procedure shown in Lines 12-23 of Algorithm \ref{alg:refinement} for mesh curving.
        \EndFor
        \EndWhile

       \end{algorithmic}
    \label{alg:coarsening}
\end{algorithm}

\begin{figure}
\centering
  \includegraphics[width=0.98\textwidth]{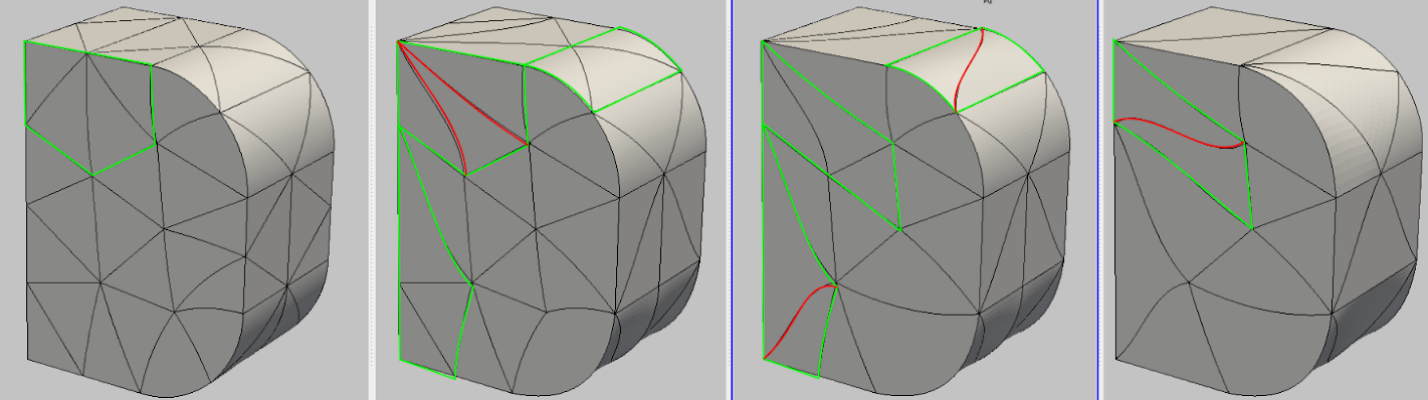}
  \caption{Demonstration of multiple iterations of coarsening for a curved mesh.}
  \label{fig:kovacrv3d}
\end{figure}

A demonstration of three iterations of coarsening on a curved mesh is shown in Figure~\ref{fig:kovacrv3d}, where portions of the initial cavity boundary are highlighted in green and the new mesh edges created after the collapse operations are highlighted in red.

\subsection{Swap Operation}

In the swap operation, the elements adjacent to a specific edge are removed, and the cavity is re-triangulated without that edge. This operation can be applied to remove either long or short edges in metric space while simultaneously improving mesh quality within the local cavity.

\begin{figure}[htbp]
\centering
	\includegraphics[trim={0 2.5cm 0 0}, clip, width=0.55\textwidth]{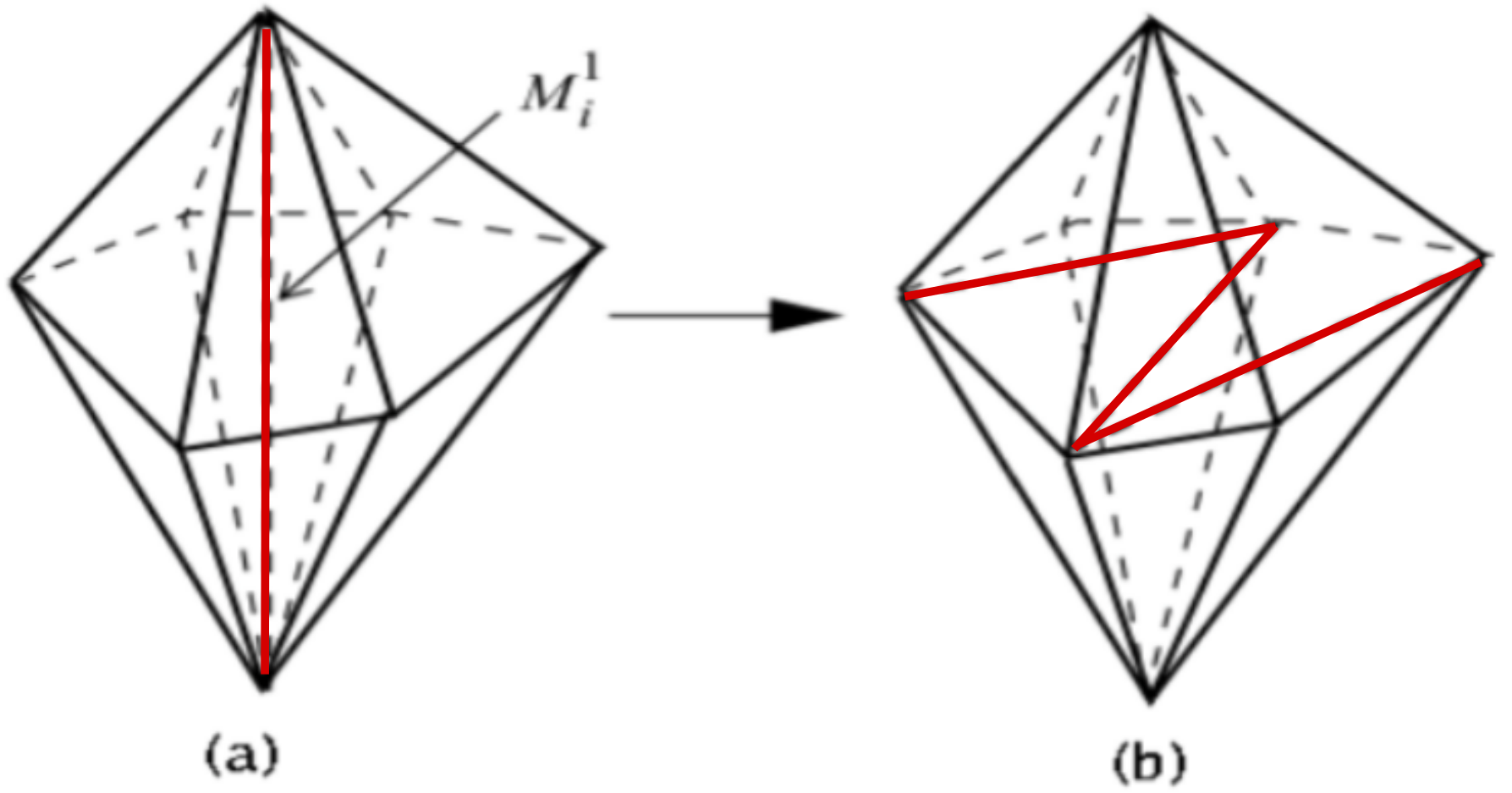}
	\caption{Mesh topology changes in a tetrahedral cavity before and after swap operation.}
	\label{fig:swap-topo}
\end{figure}

For a cavity in a $3D$ tetrahedral mesh, an output triangular mesh is created by re-triangulating the ``ring'' of vertices (those vertices opposite the central edge across an input triangle) to produce the best-quality configuration. At this stage, the best resulting cavity quality, $Q$, is computed under the assumption that the new edges and faces are straight-sided.
The number of possible topological configurations in the cavity is determined by the number of vertices in the ring, which is dictated by the number of elements adjacent to the key edge \cite{Li2005}. An example of the topological changes in the interior of a cavity during an edge swap is shown in Figure~\ref{fig:swap-topo}.

The new entities on the model boundary are curved to approximate the geometric model. When the swap operation is performed within a cavity of a curved mesh, the interior entities are curved as required to address validity and quality issues using the procedure presented in Section~\ref{InteriorCurving}.

\begin{figure}[htbp]
\centering
\fbox{\includegraphics[width=0.35\textwidth]{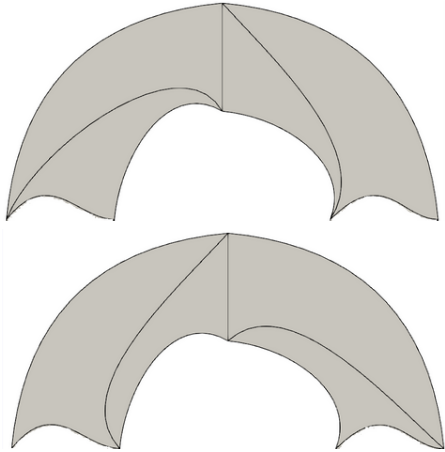}}
	\caption{Demonstration of swap operation for cubic curved mesh.}
	\label{fig:swap-annulus}
\end{figure}

Figure\ \ref{fig:swap-annulus} shows a simple example of swap operation performed on a curved cubic mesh.
In this example, the swap operation is improving element shapes by eliminating small angles.

An overview of the swap operation is given in Algorithm \ref{alg:swap}. 
\begin{algorithm}
  \caption{Swap operation}
  \begin{algorithmic}[1]

      \State INPUT: Mesh and metric field attributed with target edge lengths.
      \State OUTPUT: Cavity with modified topology and geometry.\bigskip

        \State Compare current edge lengths with target lengths and mark undesirable edges.
        \While{Edge swaps remain}
        \State Form independent set of cavities including consideration of mesh partition boundaries.
        \For{Loop over cavities}
            \State Decide topology configuration for cavities resulting in optimal quality after swap based on re-triangulation of the ring of vertices.
            \State Create topology and classification for new entities using the old cavity.
            \State Execute procedure shown in Lines 9-23 of Algorithm \ref{alg:refinement} for mesh curving.
        \EndFor
        \EndWhile

       \end{algorithmic}
    \label{alg:swap}
\end{algorithm}

%% file: ElementShapeImprove.tex
\section{Improving Element Shapes} \label{ImproveShape}

\begin{figure}[htbp]
\centering
	\fbox{\includegraphics[width=0.8\textwidth]{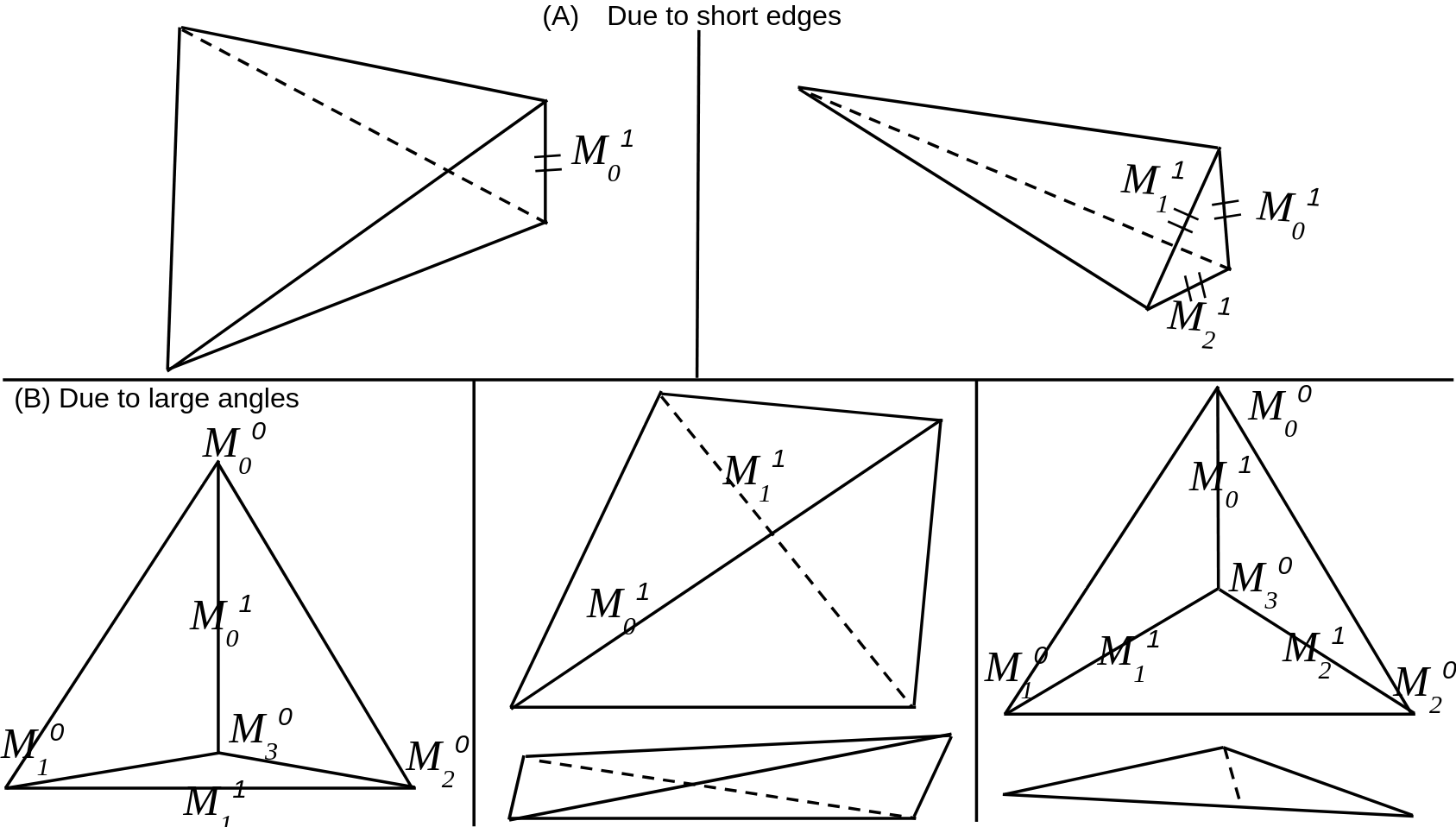}}
	\caption{Types of poor quality tetrahedra due to (A) short edges or (B) large angles.}
	\label{fig:tettypes}
\end{figure}

The application of the mesh modification operators discussed in Section \ref{ModificationOperators} focuses on meeting the given mesh metric field. There are times when their application can produce poorly shaped elements. The fix element shape procedure is focused on the targeted application of the core mesh modification procedures, plus three compound operators discussed below that specifically eliminate the poorly shaped elements. The poor-quality elements are due to either very short edges or very large dihedral angles that create a flat tetrahedron. Figure~\ref{fig:tettypes} shows the specific cases of poorly shaped tetrahedra that can arise. A key aspect of effectively eliminating the poorly shaped tetrahedra is to apply mesh modification operations based on whether the element falls into the short-edge or flat-tetrahedron case \cite{Li2003a, Wan2006} (see Algorithm~\ref{alg:fixshape}).

Information about mesh topology and geometry is used to analyze the poorly shaped elements and determine the most effective operation for execution of shape quality improvement.
The algorithm for element quality improvement is shown in Algorithm \ref{alg:fixshape}.
\begin{algorithm}
\caption{Fix Element Shape and selection of modification operation}
\begin{algorithmic}[1]

    \While {Undesired element exists and the number of low quality element decreases}
            \For {Each undesired quality element}
                \If {\textbf{Type A:}}
                    \For {Each short edge}
                        \State Collapse short edge that results in improved output quality.
                    \EndFor
                \Else {\textbf{ Type B:}}
                    \State Determine element type shown in \textbf{Figure~\ref{fig:tettypes}} based on the number of large angles.
                    \State Execute operations in the following sequence until desired element quality then break and move to next element.
                    \If {\textbf{Type B1:}}
                    \State \textbf{try and execute the first that is successful:}
                        \State \;\;\;\; Swap the long edge $M^{1}_{1}$ in the low quality triangle ($M^{0}_{1}$, $M^{0}_{2}$, $M^{0}_{3}$).
                        \State \;\;\;\; Collapse short edge connecting $M^{0}_{3}$ in the low quality triangle.
                        \State \;\;\;\; Split-collapse longest edge $M^{1}_{1}$ of low quality triangle.
                    \ElsIf {\textbf{Type B2:}}
                    \State \textbf{try and execute the first that is successful:}
                        \State \;\;\;\; Swap either of two edges $M^{1}_{0}$ or $M^{1}_{1}$ that are close to each other.
                        \State \;\;\;\; Collapse one of the short edges of the element.
                        \State \;\;\;\; Double split collapse between $M^{1}_{0}$ and $M^{1}_{1}$.
                    \ElsIf {\textbf{Type B3:}}
                    \State \textbf{try and execute the first that is successful:}
                        \State \;\;\;\; Split the base face and collapse the new edge.
                        \State \;\;\;\; Collapse edge that connect opposite vertex $M^{0}_{3}$ to the base face.
                        \State \;\;\;\; Swap an edge of the base face.
                    \EndIf
                \EndIf
            \EndFor
    \EndWhile
\end{algorithmic}
\label{alg:fixshape}
\end{algorithm}

As indicated in Line 5 of Algorithm \ref{alg:fixshape}, applying an edge collapse to any of the short edges in either direction will eliminate the short-edged tetrahedron.

The most effective means of eliminating the flat tetrahedron depends on which of the three cases of flat element is being addressed. The three cases are labeled B1, B2, and B3, depending on whether the element contains one, two, or three large angles, respectively.
The types B1, B2, and B3 are shown from left to right in the second row of Figure~\ref{fig:tettypes} (the side view of B2 and B3 is shown in the two images underneath the respective tetrahedra), and the procedures to resolve them are stated in Lines 10, 15, and 20 of Algorithm~\ref{alg:fixshape}.
As stated in Line 9, the modification operations listed in Lines 10-25 are executed in the listed sequence until the element improves to the desired quality. Once satisfactory quality is reached, execution moves on to the next element in Line 2.

\begin{figure}[htb]
\centering
	\fbox{\includegraphics[width=0.5\textwidth]{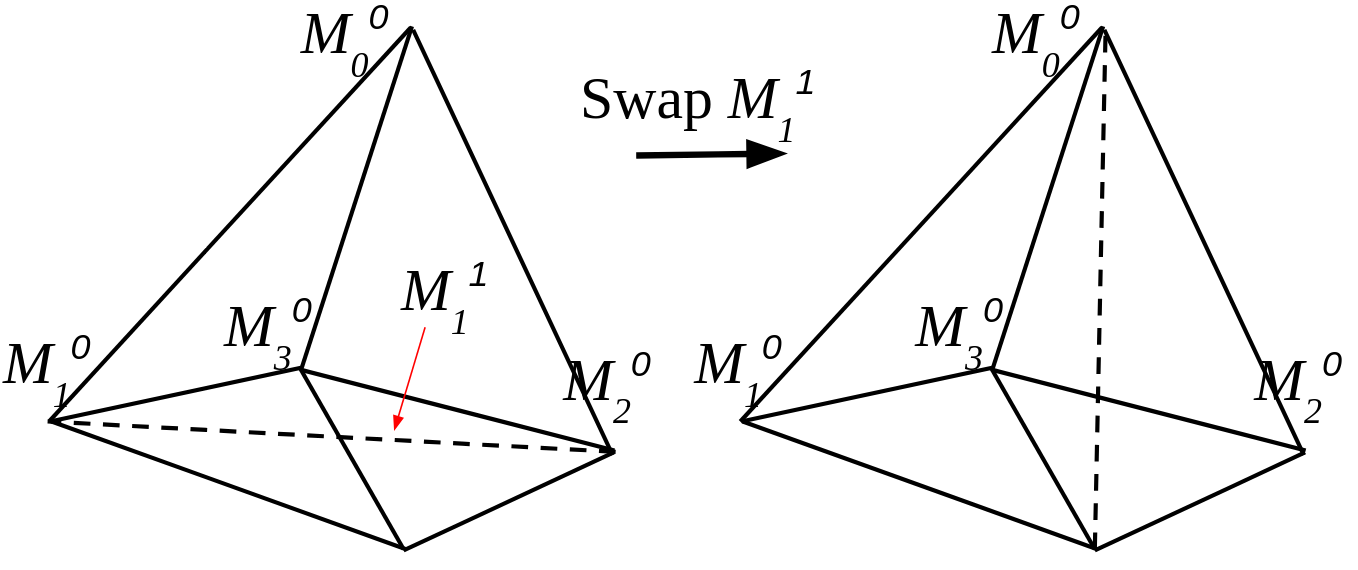}}
	\caption{Illustration of swap $M^{1}_{1}$ operation for type B1.}
	\label{fig:B1-swap}
\end{figure}

\begin{figure}[htb]
\centering
	\fbox{\includegraphics[width=0.5\textwidth]{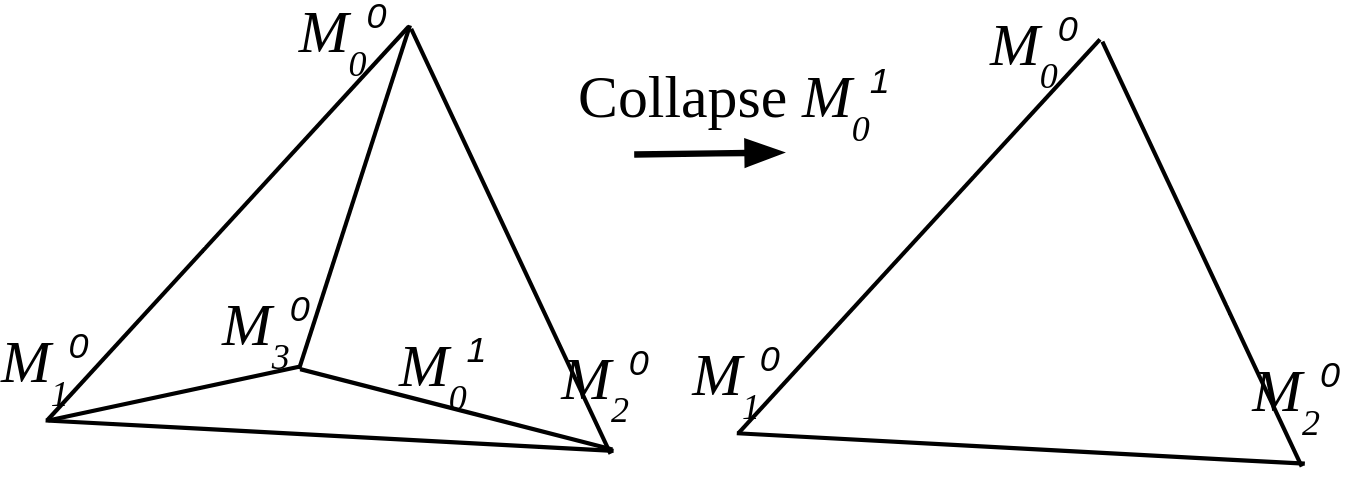}}
	\caption{Illustration of collapse operation that eliminates the flat tetrahedron of type B1.}
	\label{fig:B1-collapse}
\end{figure}

\begin{figure} [h]
\centering
	\fbox{\includegraphics[trim={0 0.5cm 0cm 0}, clip,width=0.7\textwidth]{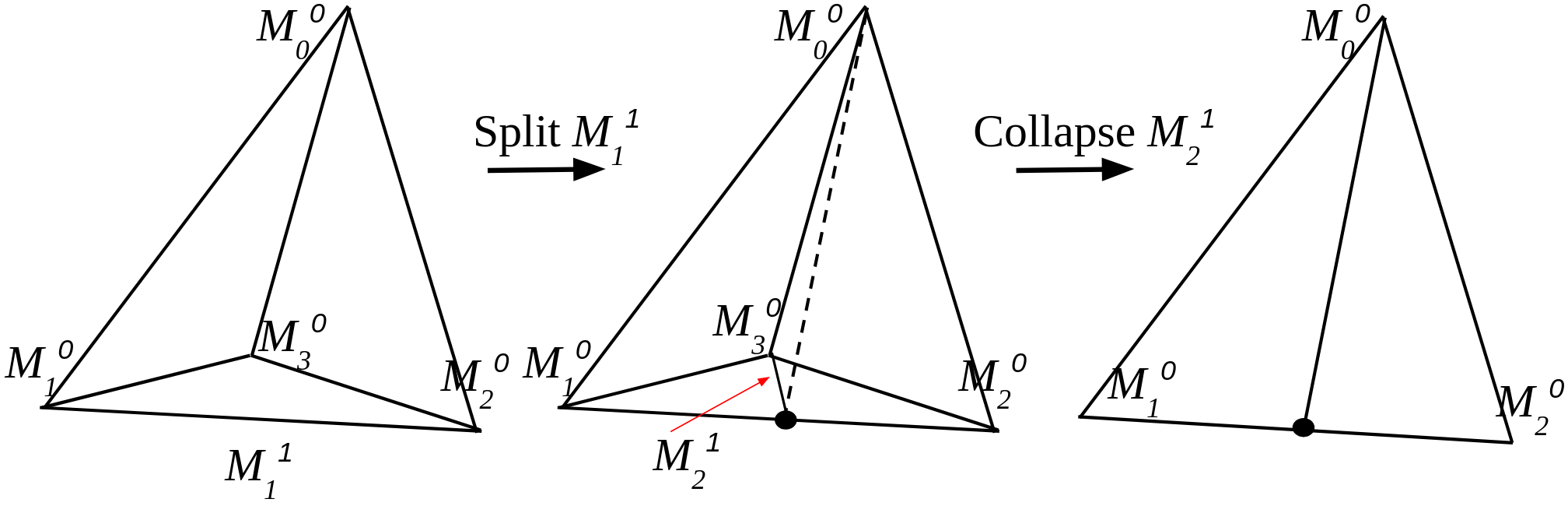}}
	\caption{Illustration of split+collapse operation on $M^{1}_{1}$ that eliminates the flat tetrahedron of type B1.}
	\label{fig:B1-splitcollapse}
\end{figure}

The execution of edge swap on long edges that are close to other entities has been found to be useful for shape improvement \cite{Li2003a}, as stated in Lines 12, 17, and 24.
Specifically, for type B1 in Line 12, the edge $M^{1}_{1}$ shown in the left image of Figure~\ref{fig:B1-swap} is a candidate for swapping.
In Line 13, edge $M^{1}_{0}$ shown in the left image of Figure~\ref{fig:B1-collapse} is collapsed to eliminate the flat tetrahedron.
In Line 14, a compound operation is considered, where edge $M^{1}_{1}$ is split followed by a collapse of $M^{1}_{2}$ (see Figure~\ref{fig:B1-splitcollapse}).

For type B2, the first operation considered is swap (Line 17). If swap is not successful, the collapse operation is considered (Line 18).

\begin{figure}[htb]
\centering
	\fbox{\includegraphics[trim={0 0cm 0cm 0.5cm}, clip,width=0.98\textwidth]{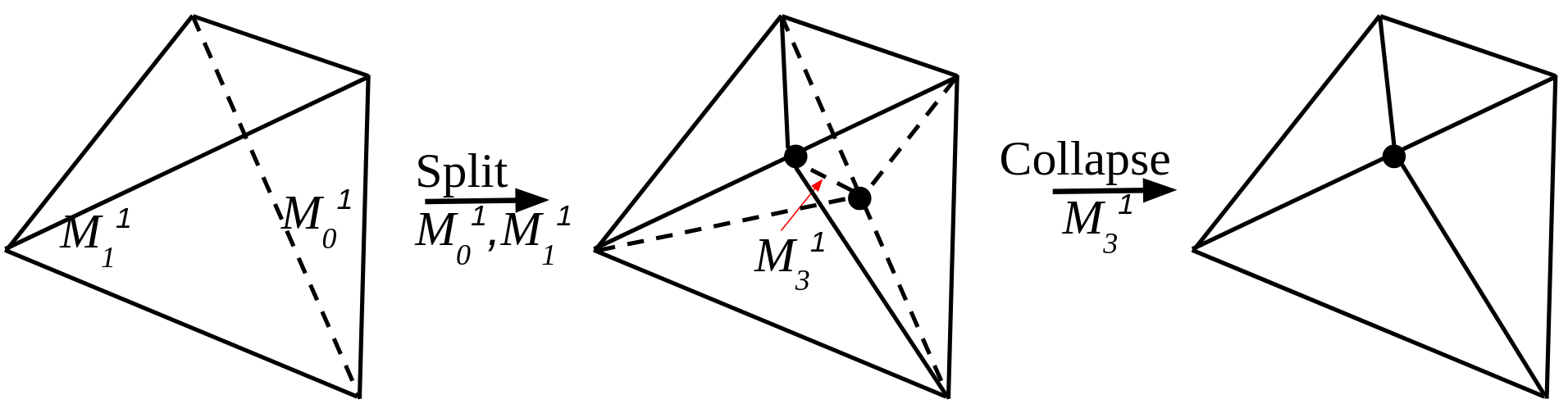}}
	\caption{Illustration of double split+collapse operation that eliminates the flat tetrahedron of type B2.}
	\label{fig:B2-dbsplitcollapse}
\end{figure}

If neither swap nor collapse is successful, the compound operation of a double split of edges $M^{1}_{0}$ and $M^{1}_{1}$, followed by the collapse of $M^{1}_{3}$, eliminates the flat tetrahedron (see Figure~\ref{fig:B2-dbsplitcollapse}).

\begin{figure}[htb]
\centering
	\fbox{\includegraphics[width=0.98\textwidth]{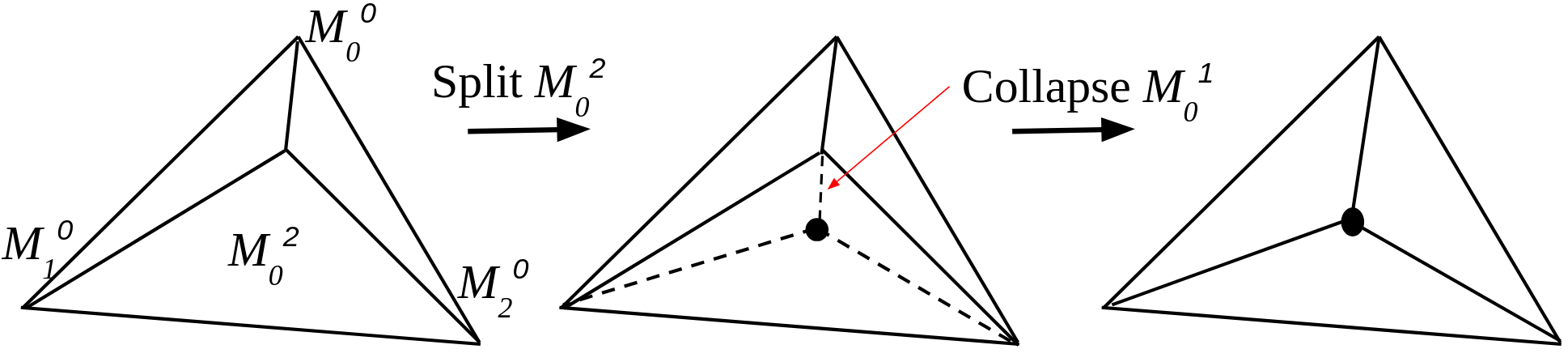}}
	\caption{Illustration for type B3: split base face $M^{2}_{0}$ and collapse the new edge $M^{1}_{0}$ to eliminate the flat tetrahedron.}\label{fig:B3-facesplitcollapse}
\end{figure}

For type B3, the first operation considered (Line 22) is the compound operation of a face $M_0^2$ split followed by the collapse of $M_0^1$ to eliminate the flat tetrahedron (see Figure~\ref{fig:B3-facesplitcollapse}).
If that is not successful, a collapse operation is considered. If neither the compound operation nor the collapse is successful, a swap operation is considered.

During the application of the above-mentioned operators, the procedures discussed in Section \ref{InteriorCurving} are employed as needed to curve the mesh geometry.


%% file: ExecuteCurvedMeshAdapt.tex
\section{Execution of Curved Mesh Adaptation} \label{ExecuteResults}

\subsection{Curved Mesh Modification Text Cases}

The desired mesh sizes in the adaptation procedure are described using a metric field \cite{compere2010mesh, Li2005, loseille2015parallel}. 
The input target metric field is provided at mesh vertices as the desired lengths of adjacent mesh edges. The following are a set of test cases demonstrating the application of the mesh modification procedures.

\begin{figure}
\captionsetup{width=\linewidth, justification=centering}
    \centering
  \fbox{\includegraphics[width=0.6\textwidth]{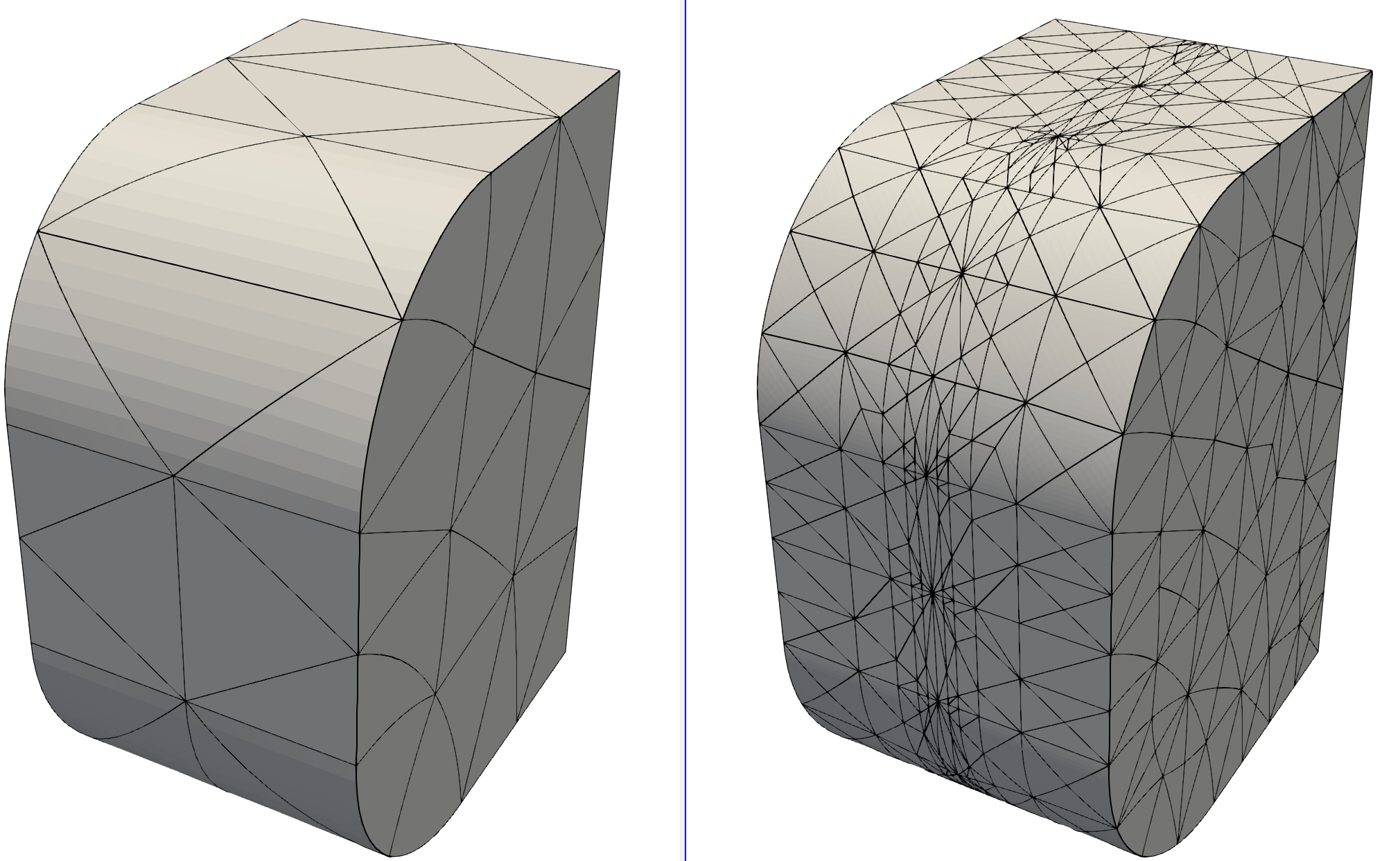}}
  \caption{Demonstration of cubic mesh adaptation with anisotropic size field on a convex geometry.}
    \label{fig:kova-aniso-refine-example}
\end{figure}

Figure~\ref{fig:kova-aniso-refine-example} shows the application of curved mesh adaptation for a mildly anisotropic size field \cite{park2015comparing} described in Eq.~\ref{eq:ugawg-anisoSf}.
\begin{equation} \label{eq:ugawg-anisoSf}
h=\begin{bmatrix}
h_x \\
h_y \\
h_z \\
\end{bmatrix}
\quad\text{where}
\left\{\begin{tabular}{l}
$h_x = 0.1$ \\
$h_y = 0.1$ \\
$h_z = 0.001+ 0.198 \; | z - 0.5 |$ \\
\end{tabular}\right.
\end{equation}

\begin{figure}
\captionsetup{width=\linewidth, justification=centering}
    \centering
  \fbox{\includegraphics[width=0.6\textwidth]{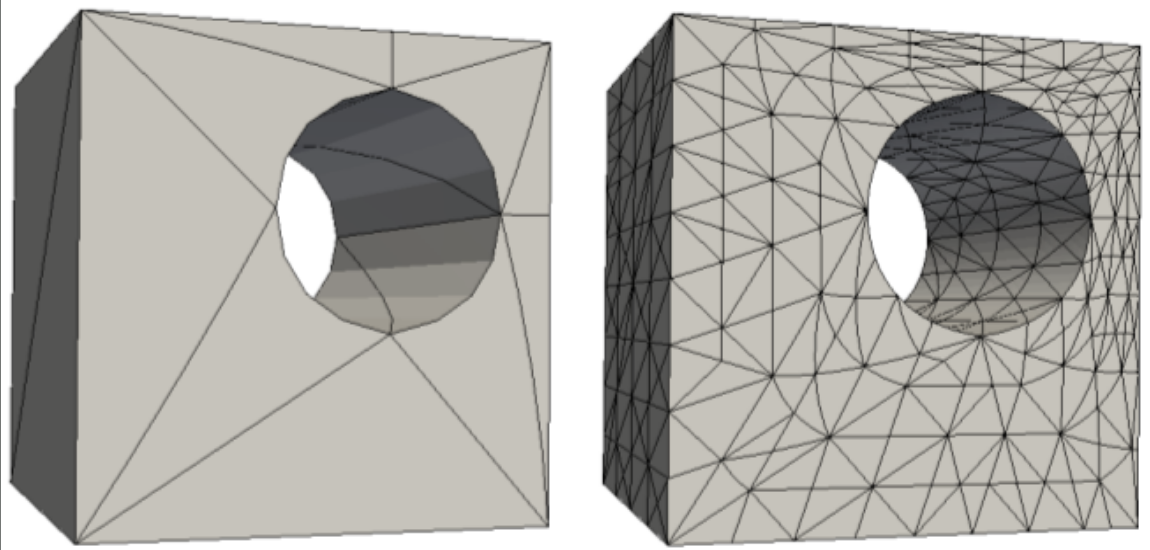}}
  \caption{Demonstration of high order adaptation with anisotropic size field that shows the ability to deal with concavity in the model: initial coarse mesh and adapted mesh of cubic geometry.}
  \label{fig:boxCirc-aniso-refine-examples}
\end{figure}

Figure~\ref{fig:boxCirc-aniso-refine-examples} shows an initial and adapted cubic curved mesh of a model that demonstrates the ability of the adaptive procedure to handle concavities in the geometric model.

\begin{figure}[htbp]
    \centering
    \captionsetup{width=\linewidth, justification=centering}
    \setlength{\fboxrule}{1pt}
    \fbox{\includegraphics[width=0.58\textwidth]{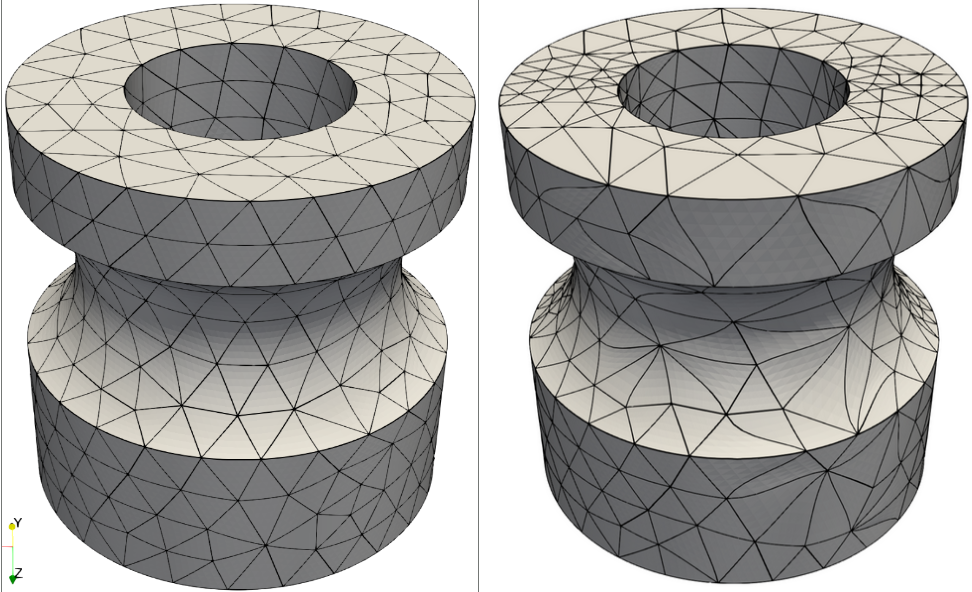}}
    \caption{Demonstration of mesh modification operators on adapted mesh with anisotropic size field.}
    \label{fig:cyl-grv-adapt}
\end{figure}

\begin{figure}[htbp]
    \centering
    \captionsetup{width=\linewidth, justification=centering}
  \fbox{\includegraphics[width=0.55\textwidth]{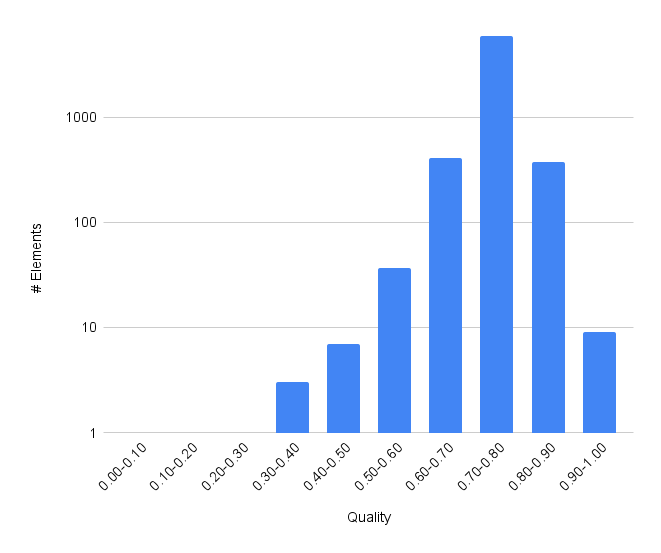}}
  \caption{Figure shows histogram of mesh quality of the adapted mesh for annular cylinder case.}
  \label{fig:cylgrv-qual}
\end{figure}

Figure \ref{fig:cyl-grv-adapt} shows the demonstration of mesh modification operators on initial mesh adapted to a target anisotropic metric field.
The annulus cylindrical model has a concavity in the form of an elliptical groove cut out from the volume.
Figure \ref{fig:cylgrv-qual} shows the histogram of the mesh quality for the adapted curved mesh with the worst element quality of 0.191.

\begin{figure}
  \fbox{\includegraphics[width=0.98\linewidth]{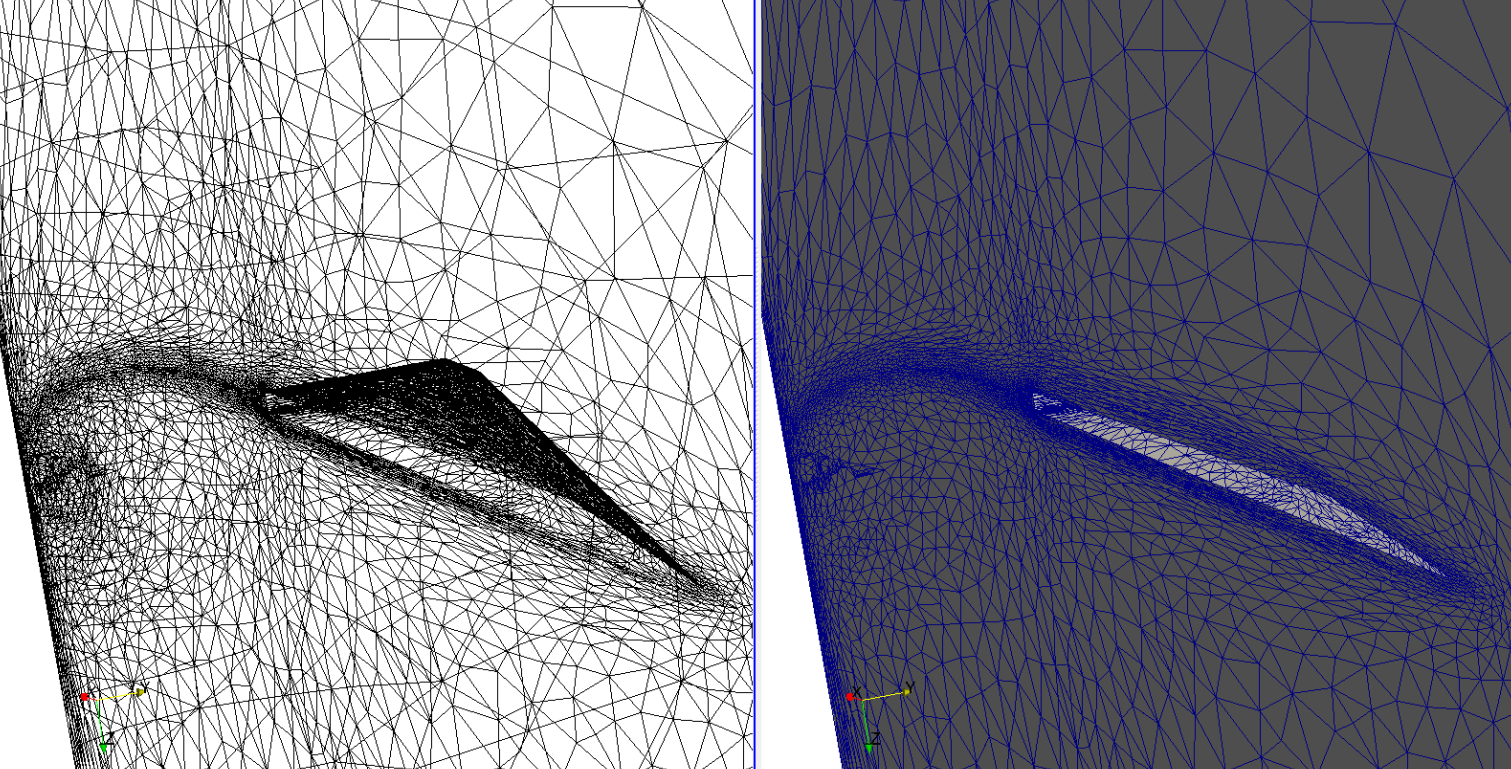}}
  \caption{Mesh of the delta wing in wireframe view (left) and surface view (right) containing roughly 550 thousand elements.}
  \label{fig:wing}
\end{figure}

\begin{figure}
  \centering
  \fbox{\includegraphics[width=0.55\linewidth]{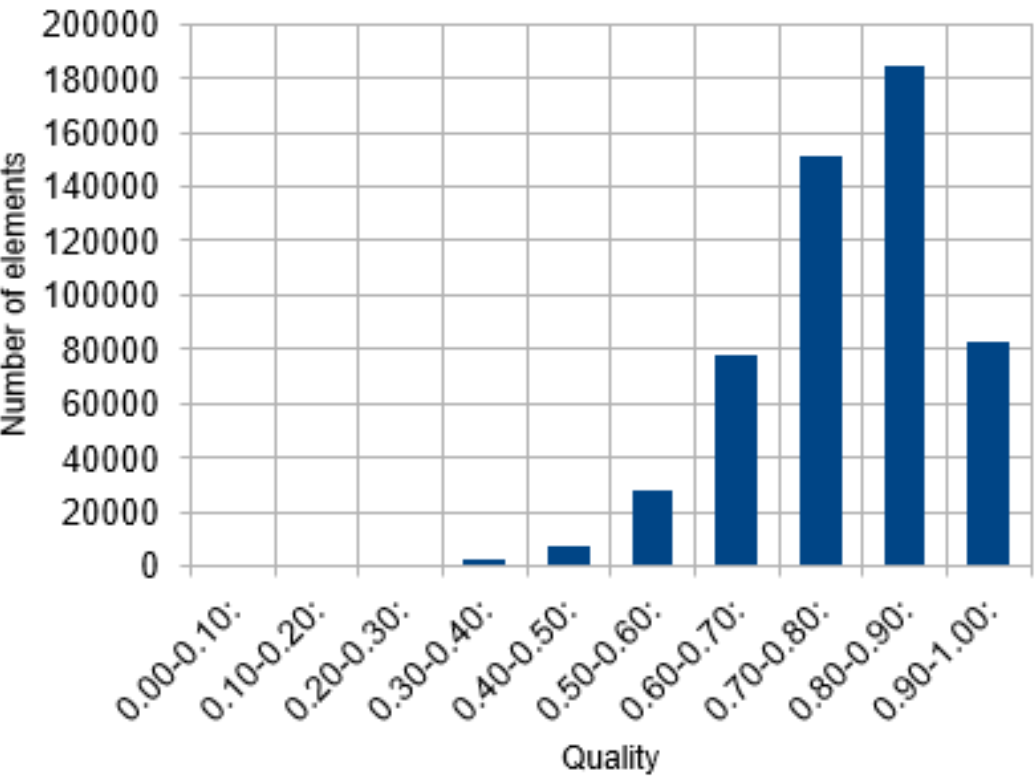}}
  \caption{Distribution of element qualities after anisotropic adaptation for delta wing case.}
  \label{fig:50qual}
\end{figure}

A demonstration of mesh adaptation for the Delta wing example \cite{ugawg} is shown in Figure~\ref{fig:wing}, that shows a close up near the wing of the adapted mesh. 
Figure \ref{fig:50qual} shows distribution of element qualities in the adapted mesh.
Table \ref{tab:deltawing} indicates the mesh adaptation execution times for two examples on the Delta wing model, where the second case has a finer target mesh.
\begin{table}
\captionsetup{width=\linewidth, justification=centering}
\caption{Performance of mesh adaptation for Delta wing adaptive example.}
\centering
\begin{tabular}{|c|c|c|}
\hline
 Final entity count & Computational resources & Time(s)\\[0.5ex]
 \hline\hline
  region=534,941; edge=631,875 & 1 node, 1CPU+1GPU(A100) & 4.79\\[1ex]
 \hline
   region=5,284,180; edge=6,216,939 & 1 node, 1CPU+1GPU(A100) & 48.23\\[1ex]
 \hline
\end{tabular}
\label{tab:deltawing}
\end{table}


\begin{figure}
\captionsetup{width=\linewidth, justification=centering}
    \centering
  \includegraphics[width=0.55\textwidth,  trim={0cm 0cm 0cm 2cm}]{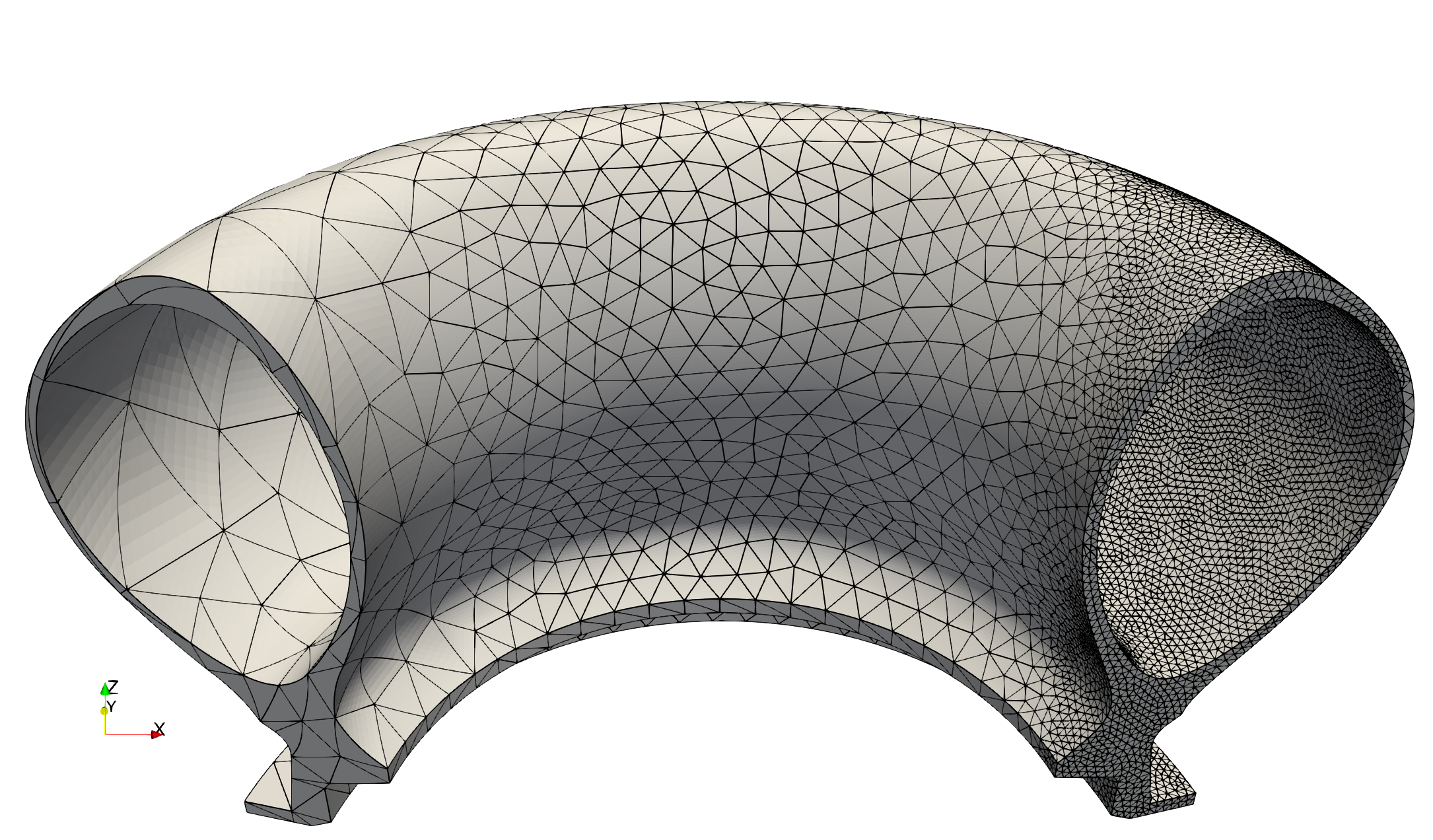}
  \caption{Curved mesh for the edge plasma region of C-Mod tokamak that demonstrates the application of isotropic spatially varying size field.}
  \label{fig:cmod-elecDens}
\end{figure}

For fusion physics analysis, the edge plasma region is defined between the antenna and the core plasma region.
Figure~\ref{fig:cmod-elecDens} demonstrates the application of a varying isotropic size field to generate a graded curved mesh in the edge region of the C-Mod tokamak.

\subsection{Adaptive Simulation Results}

The driving application for the curved mesh adaptation procedures presented in this paper is electromagnetic (EM) radio-frequency (RF) analysis of magnetically confined fusion systems using the Petra-M \cite{shiraiwa2017} / MFEM \cite{mfem} simulation framework.
Petra-M is an EM simulation tool for modeling RF full-wave propagation in plasmas. Petra-M uses the MFEM open-source scalable C++ finite element library and has been used for full-wave simulations. The example adaptive simulation result presented below is the full-wave RF simulation of a J-port RF antenna in the Alcator C-Mod tokamak~\cite{diab2026comparison}.

\begin{figure}[htbp]
\centering
\captionsetup{width=\textwidth}
	\includegraphics[width=0.68\textwidth]{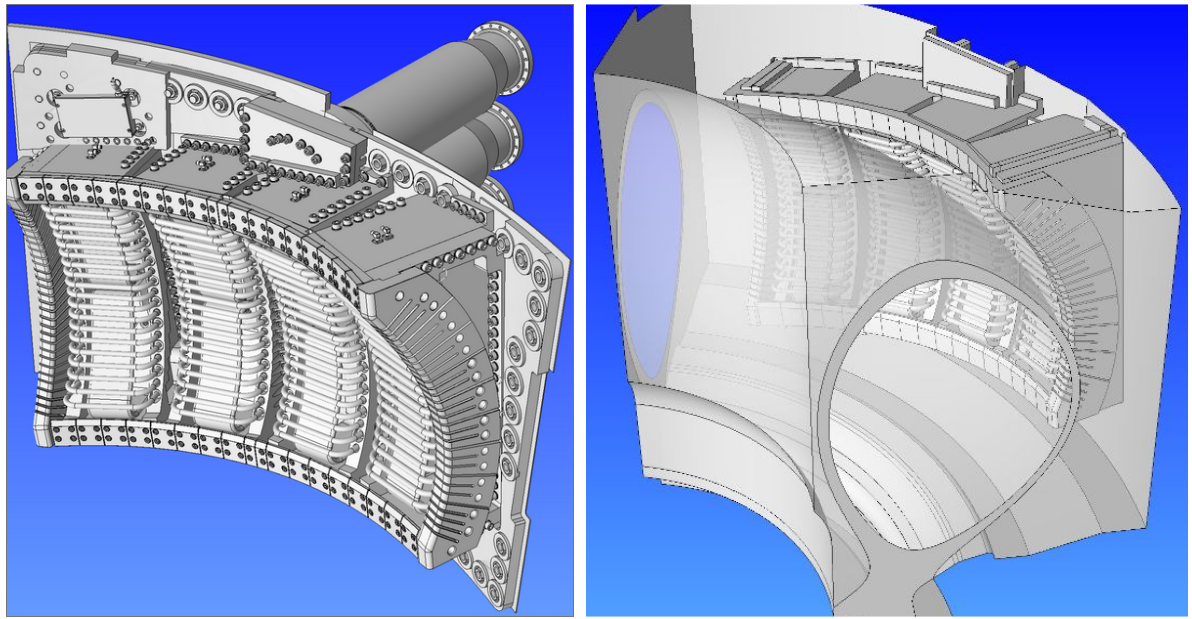}
	\caption{Initial manufacturing model of CMOD RF antenna and non-manifold analysis model included in the tokamak.}
	\label{fig:cmod-model}
\end{figure}

There are three main steps in the execution of an adaptive Petra-M/MFEM RF simulation of a fusion system. The first step is the construction of the analysis CAD model. This process includes combining cleaned-up and de-featured antenna CAD models with a CAD model that includes the wall geometry and selected physics geometry in the form of key flux surfaces~\cite{diab2026comparison, shephard2024unstructured}. Figure~\ref{fig:cmod-model} shows the original antenna CAD model and the final assembled analysis CAD model for the demonstration adaptive simulation.

\begin{figure}
    \centering
    \captionsetup{width=\linewidth, justification=centering}
    \includegraphics[width=0.45\textwidth]{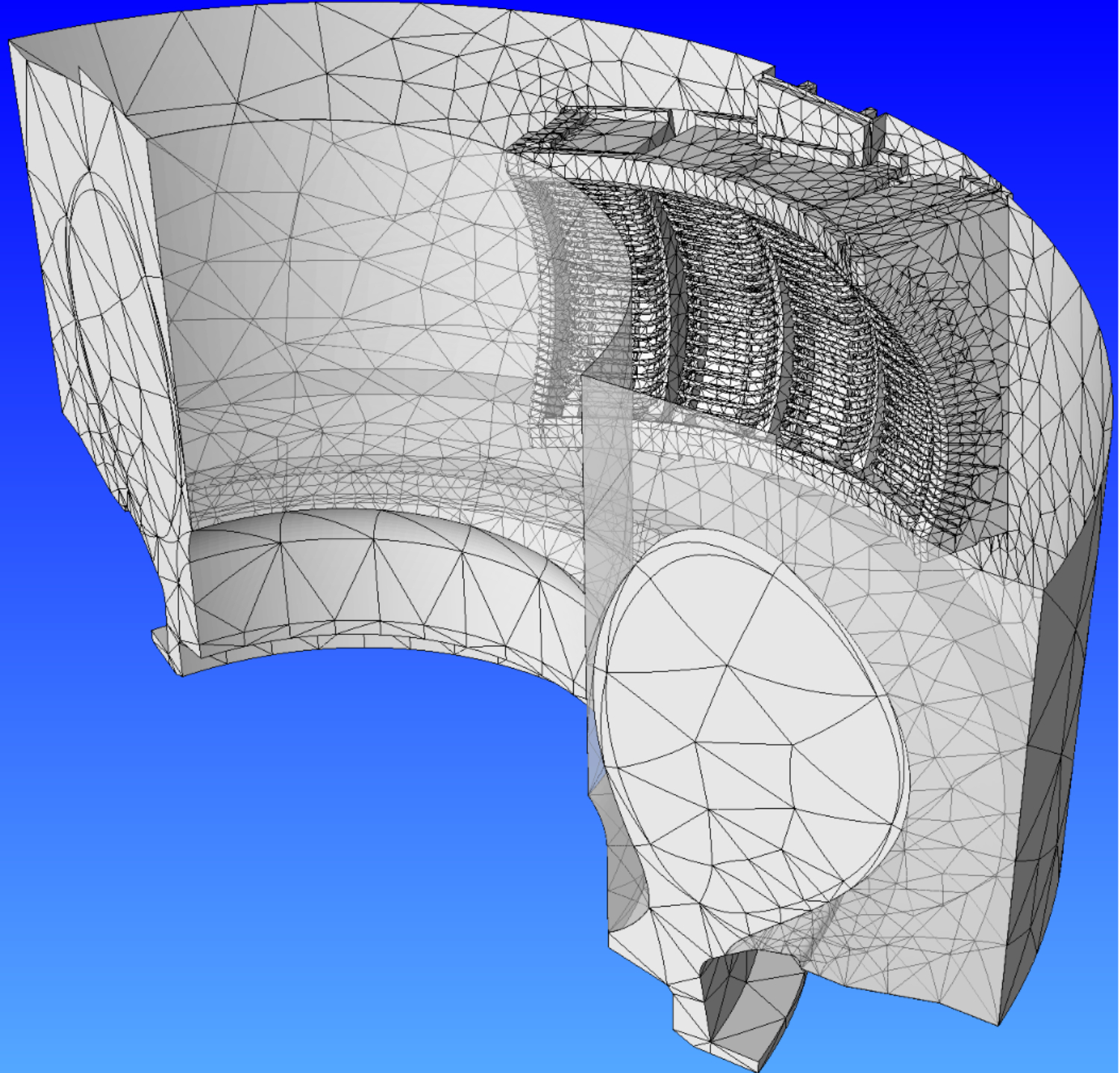}
    \caption{A fifth order curved initial mesh used for RF simulations.}
    \label{fig:RF-mesh-copy}
\end{figure}

The second step is the setup of the Petra-M/MFEM problem specification and input. Given the analysis material properties, boundary conditions, and loads defined on entities in the CAD model, this information can be properly transferred to the initial mesh and the adaptive meshes as they are constructed. The initial mesh is generated by first automatically producing a quadratically curved mesh \cite{simmetrix_web}, followed by inflating the order of geometric approximation for mesh edges and faces classified on curved boundaries using the procedure given in Section~\ref{CurvingToBoundary}. Figure~\ref{fig:RF-mesh-copy} shows views of a fifth-order coarse initial mesh used for the adaptive tokamak RF simulation.

The third step is the execution of the adaptive simulation loop, which is performed by Petra-M/MFEM and the in-memory integration of the curved mesh modification procedures described in this paper. Reference~\cite{siboni2022adaptive} describes the procedure used to perform error estimation and compute the new mesh size field.


The propagation of RF electromagnetic waves is modeled by the time-harmonic Maxwell’s equations \cite{pMonk} and the cold plasma approximation in the plasma core.
The perfect electric conductor (PEC) boundary condition is applied on the surfaces of the limiters, antenna straps, and associated feeding structures. Interior flux surfaces are specified as continuity boundaries, and RF power is injected into the vacuum vessel through the cylindrical coaxial ports at the back of the antenna by application of the Port boundary condition~\cite{diab2026comparison}. The problem is solved using fifth-order Nédélec finite elements.

\begin{figure}
    \centering
    \captionsetup{width=\linewidth, justification=centering}
    \setlength{\fboxrule}{1pt}\fbox{\includegraphics[trim={0.4cm 0 0.5cm 0}, clip, width=0.68\textwidth]{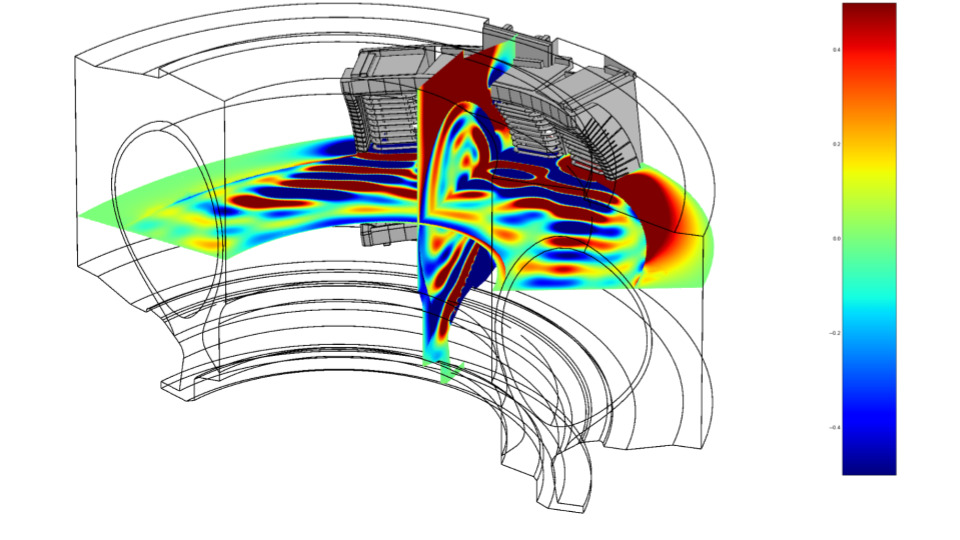}}
    \caption{Electric field obtained on an adapted mesh of fifth order elements.}
    \label{fig:RF-adapt}
\end{figure}

\begin{figure}[htbp]
    \centering
    \captionsetup{width=\linewidth, justification=centering}
    \setlength{\fboxrule}{1pt}
    \fbox{\includegraphics[width=0.78\textwidth]{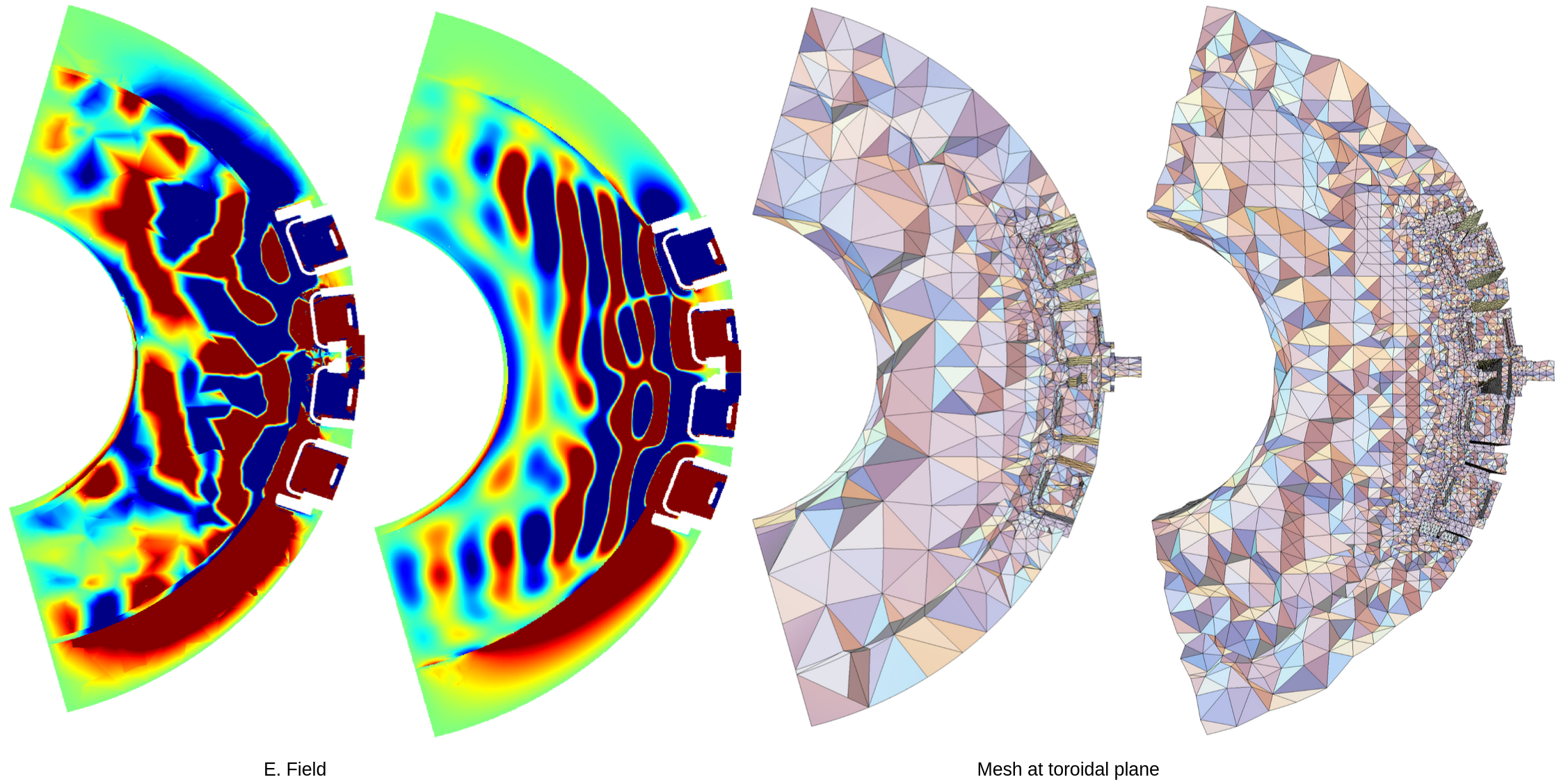}}
    \caption{Electric field and mesh on toroidal plane for initial and adapted mesh of CMOD tokamak and RF antenna.}
    \label{fig:RF-adapt-tor}
\end{figure}

\begin{figure}[htbp]
    \centering
    \captionsetup{width=\linewidth, justification=centering}
    \setlength{\fboxrule}{1pt}
    \fbox{\includegraphics[width=0.68\textwidth]{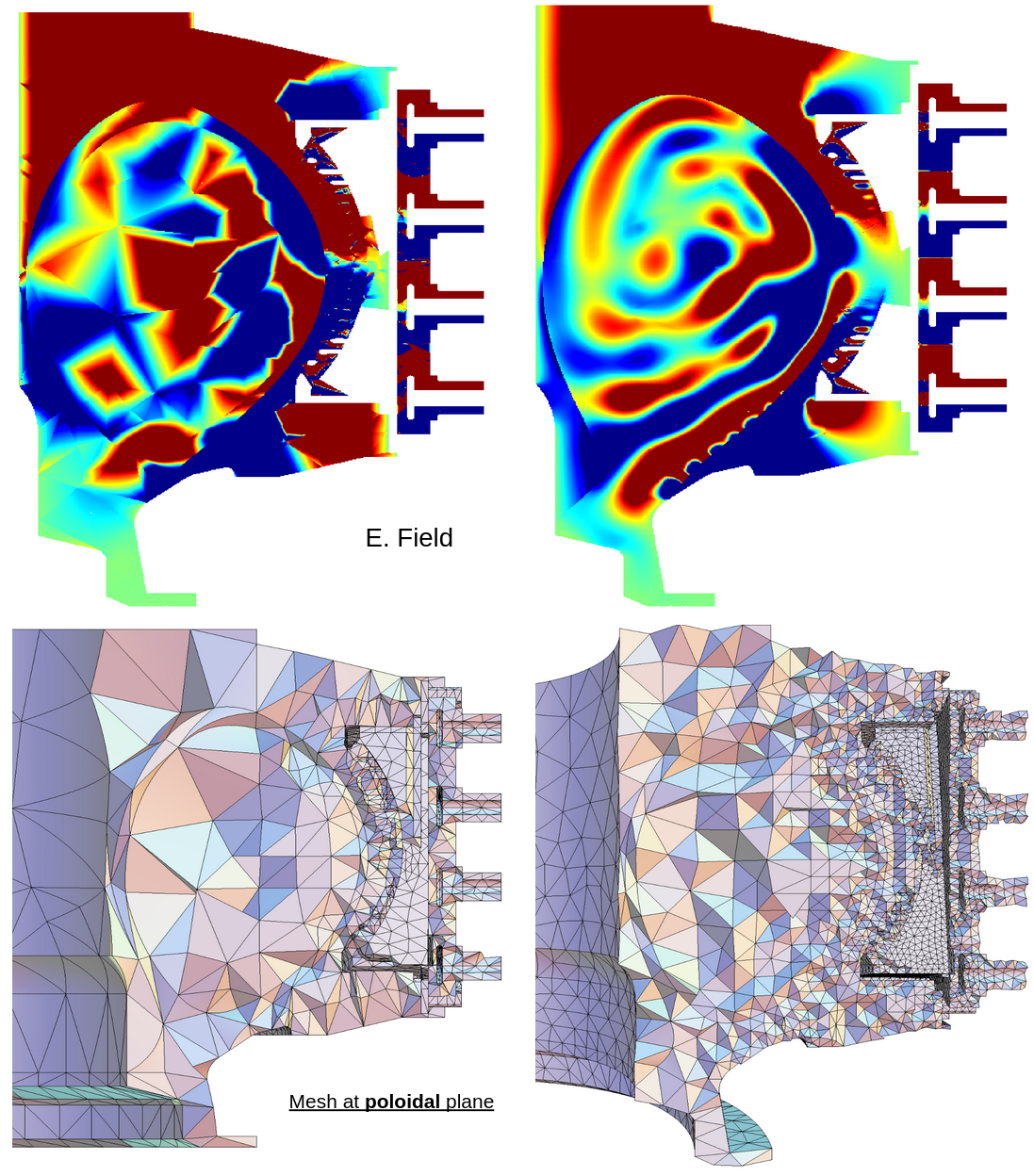}}
    \caption{Electric field and mesh on poloidal plane for initial and adapted mesh.}
    \label{fig:RF-adapt-pol}
\end{figure}

\begin{figure}
    \centering
    \captionsetup{width=\linewidth, justification=centering}
    \includegraphics[width=0.75\textwidth]{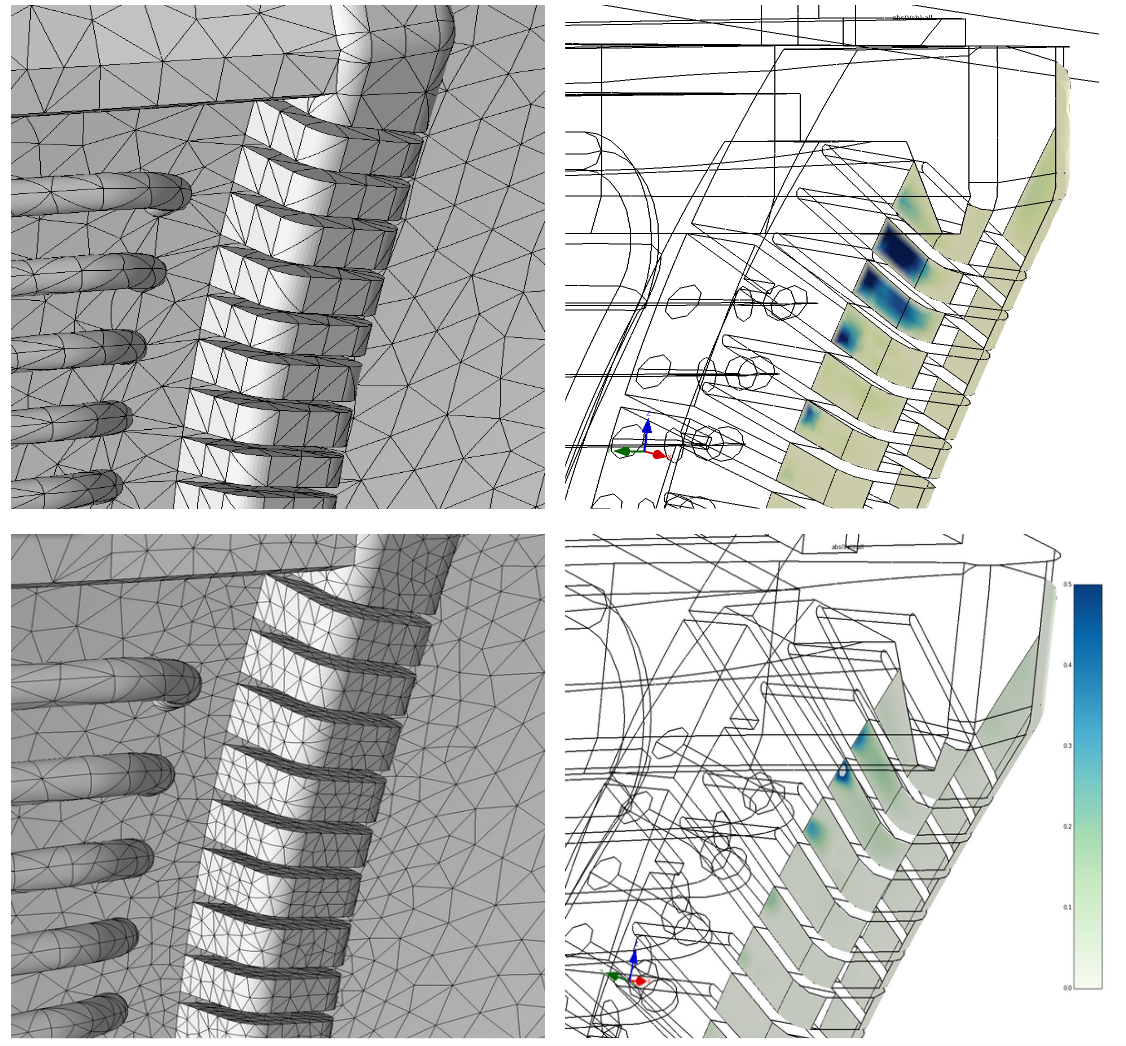}
    \caption{Close-up near antenna limiter and faraday grid: Initial and adapted meshes in left column shown with RF voltage in the right column.}
    \label{fig:RF-adapt-volt}
\end{figure}

\begin{figure}[htbp]
    \centering
    \captionsetup{width=\linewidth, justification=centering}
  \fbox{\includegraphics[width=0.75\textwidth]{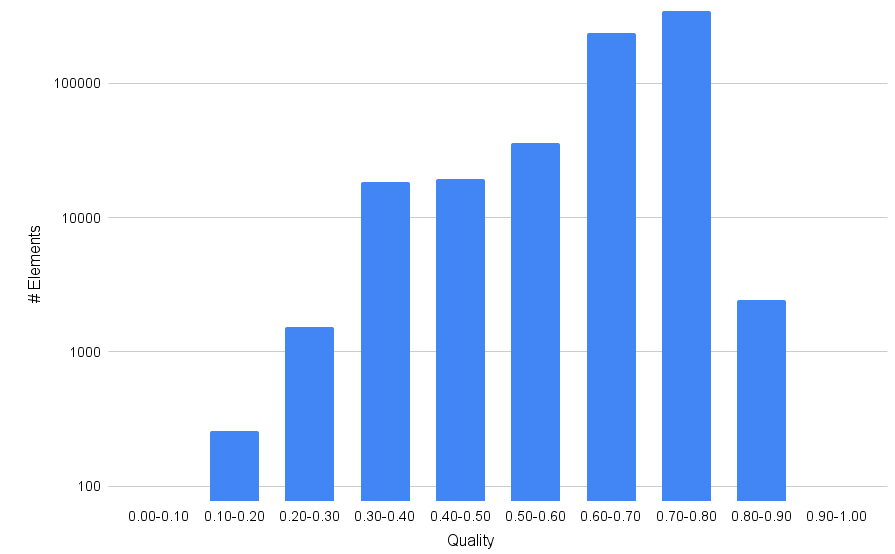}}
  \caption{Figure shows histogram of mesh quality of the adapted mesh for CMOD and RF antenna case.}
  \label{fig:cmodQual-semilog}
\end{figure}

Figure~\ref{fig:RF-adapt} shows the RF antenna along with the electric field obtained on the adapted mesh composed of fifth-order elements, using the mesh shown in Figure~\ref{fig:RF-mesh-copy} as the starting mesh.
Figures~\ref{fig:RF-adapt-tor} and~\ref{fig:RF-adapt-pol} show the electric field and mesh on the respective toroidal and poloidal center planes. Figure~\ref{fig:RF-adapt-volt} shows a close-up of the initial and adapted meshes. We observe an improvement in the estimation of RF voltage on the adapted mesh.
Figure~\ref{fig:cmodQual-semilog} shows the histogram of mesh quality for the adapted curved mesh, with the worst element quality being 0.146. 

Table \ref{tab:cmod-orders} provides the numbers of curved elements in the initial and adapted meshes. In the initial mesh 31\% of the elements were curved to fifth order because they had a mesh face or mesh edge classified on a curved model boundary. In the adapted mesh this was reduced to 15\%. In the initial mesh 2.5\% of the interior elements were curved to cubic order while in the adapted mesh this was 1.0\%. This means that in the initial mesh 66\% of the elements were straight straight-sided and planar-faced, while in the adapted mesh 84\% of the elements were straight straight-sided and planar-faced. 

The execution time required for the mesh adaptation process was 94.5 seconds while the solution of the adapted mesh took 849.4 seconds. 

\begin{table}[htbp]
\captionsetup{width=\linewidth, justification=centering}
\caption{Element orders for mesh entities on boundary and interior.}
\centering
\begin{tabular}{|c|c|c|c|}
\hline
   & Total elements & Order 5 boundary elements & Cubic interior elements \\[1ex]
  \hline\hline
   Initial mesh & 107,722 & 33,576 & 2,681 \\[1ex]  \hline
   Adapted mesh & 643,065 & 97,251 & 6,738 \\[1ex]  \hline
\end{tabular}
\label{tab:cmod-orders}
\end{table}

%% file: Closing.tex
\section{Closing Remarks}\label{Closing} 

The increased use of high-order finite element methods requires the ability to generate and adapt
curved geometric elements that accurately represent curved model geometry at the level of
approximation required to ensure solution accuracy. This paper presents a set of extensions to
previously developed curved mesh generation and adaptation procedures, aimed at more
effectively addressing the needs of high-order finite element simulations on general geometries.

One area of development concerns the generation of curved mesh edges and faces that accurately
represent curved model boundaries. By exploiting the properties of B\'ezier mesh geometry and
robust cavity-based mesh modification procedures, it is possible to construct valid meshes that
approximate curved model boundaries to any desired order of accuracy.

In the case of the target application involving vector finite elements, oscillations in the interpolated mesh geometry, even when
using optimal interpolation point distributions, led to
reduced accuracy in quantities of interest associated with the domain boundary. To address this
issue, it was found that using an optimized B\'ezier-based geometric approximation of the model
surface eliminates these oscillations and restores the desired accuracy. Further investigation is
required to determine the level and type of geometric approximation needed for accurate prediction
of a broader range of quantities of interest.

To avoid the cost of generating and maintaining high-order mesh geometry throughout the entire
domain, the curved mesh adaptation procedures developed here reduce the geometric order of
interior mesh entities to the lowest possible level while maintaining acceptable element shape
quality. In these procedures, each interior mesh entity (edges, faces, and regions) has its geometric
order explicitly specified, and its geometry is maintained accordingly.

It was found that limiting interior mesh geometry to cubic order is sufficient to provide the flexibility
needed to construct valid curved elements without requiring additional mesh refinement beyond
that dictated by the adaptive procedure. An additional advantage of this restriction is that it enables
the development of efficient procedures for constructing cubic mesh edges and faces.